\pdfoutput=1
\documentclass[11pt,twoside,a4paper,cmspaper,final,collab]{cms-tdr}
\def\svnVersion{32942}\def\cmsCernNo{2010-050}\def\cmsCernDate{\today}\def\cmsMessage{Submitted to the Journal of High Energy Physics}

\begin{document}\cmsNoteHeader{EWK-10-002}
\hyphenation{env-iron-men-tal}
\hyphenation{had-ron-i-za-tion}
\hyphenation{cal-or-i-me-ter}
\hyphenation{de-vices}
\RCS$Revision: 32641 $
\RCS$HeadURL: svn+ssh://alverson@svn.cern.ch/reps/tdr2/papers/EWK-10-002/trunk/EWK-10-002.tex $
\RCS$Id: EWK-10-002.tex 32641 2011-01-19 07:44:52Z ghm $
%
%
%

\providecommand {\etal}{\mbox{et al.}\xspace} 
\providecommand {\ie}{\mbox{i.e.}\xspace}     
\providecommand {\eg}{\mbox{e.g.}\xspace}     
\providecommand {\etc}{\mbox{etc.}\xspace}     
\providecommand {\vs}{\mbox{\sl vs.}\xspace}      
\providecommand {\mdash}{\ensuremath{\mathrm{-}}} 

\providecommand {\Lone}{Level-1\xspace} 
\providecommand {\Ltwo}{Level-2\xspace}
\providecommand {\Lthree}{Level-3\xspace}

\providecommand{\ACERMC} {\textsc{AcerMC}\xspace}
\providecommand{\ALPGEN} {{\textsc{alpgen}}\xspace}
\providecommand{\CHARYBDIS} {{\textsc{charybdis}}\xspace}
\providecommand{\CMKIN} {\textsc{cmkin}\xspace}
\providecommand{\CMSIM} {{\textsc{cmsim}}\xspace}
\providecommand{\CMSSW} {{\textsc{cmssw}}\xspace}
\providecommand{\COBRA} {{\textsc{cobra}}\xspace}
\providecommand{\COCOA} {{\textsc{cocoa}}\xspace}
\providecommand{\COMPHEP} {\textsc{CompHEP}\xspace}
\providecommand{\EVTGEN} {{\textsc{evtgen}}\xspace}
\providecommand{\FAMOS} {{\textsc{famos}}\xspace}
\providecommand{\GARCON} {\textsc{garcon}\xspace}
\providecommand{\GARFIELD} {{\textsc{garfield}}\xspace}
\providecommand{\GEANE} {{\textsc{geane}}\xspace}
\providecommand{\GEANTfour} {{\textsc{geant4}}\xspace}
\providecommand{\GEANTthree} {{\textsc{geant3}}\xspace}
\providecommand{\GEANT} {{\textsc{geant}}\xspace}
\providecommand{\HDECAY} {\textsc{hdecay}\xspace}
\providecommand{\HERWIG} {{\textsc{herwig}}\xspace}
\providecommand{\HIGLU} {{\textsc{higlu}}\xspace}
\providecommand{\HIJING} {{\textsc{hijing}}\xspace}
\providecommand{\IGUANA} {\textsc{iguana}\xspace}
\providecommand{\ISAJET} {{\textsc{isajet}}\xspace}
\providecommand{\ISAPYTHIA} {{\textsc{isapythia}}\xspace}
\providecommand{\ISASUGRA} {{\textsc{isasugra}}\xspace}
\providecommand{\ISASUSY} {{\textsc{isasusy}}\xspace}
\providecommand{\ISAWIG} {{\textsc{isawig}}\xspace}
\providecommand{\MADGRAPH} {\textsc{MadGraph}\xspace}
\providecommand{\MCATNLO} {\textsc{mc@nlo}\xspace}
\providecommand{\MCFM} {\textsc{mcfm}\xspace}
\providecommand{\MILLEPEDE} {{\textsc{millepede}}\xspace}
\providecommand{\ORCA} {{\textsc{orca}}\xspace}
\providecommand{\OSCAR} {{\textsc{oscar}}\xspace}
\providecommand{\PHOTOS} {\textsc{photos}\xspace}
\providecommand{\PROSPINO} {\textsc{prospino}\xspace}
\providecommand{\PYTHIA} {{\textsc{pythia}}\xspace}
\providecommand{\SHERPA} {{\textsc{sherpa}}\xspace}
\providecommand{\TAUOLA} {\textsc{tauola}\xspace}
\providecommand{\TOPREX} {\textsc{TopReX}\xspace}
\providecommand{\XDAQ} {{\textsc{xdaq}}\xspace}

\providecommand {\DZERO}{D\O\xspace}     


\providecommand{\de}{\ensuremath{^\circ}}
\providecommand{\ten}[1]{\ensuremath{\times \text{10}^\text{#1}}}
\providecommand{\unit}[1]{\ensuremath{\text{\,#1}}\xspace}
\providecommand{\mum}{\ensuremath{\,\mu\text{m}}\xspace}
\providecommand{\micron}{\ensuremath{\,\mu\text{m}}\xspace}
\providecommand{\cm}{\ensuremath{\,\text{cm}}\xspace}
\providecommand{\mm}{\ensuremath{\,\text{mm}}\xspace}
\providecommand{\mus}{\ensuremath{\,\mu\text{s}}\xspace}
\providecommand{\keV}{\ensuremath{\,\text{ke\hspace{-.08em}V}}\xspace}
\providecommand{\MeV}{\ensuremath{\,\text{Me\hspace{-.08em}V}}\xspace}
\providecommand{\GeV}{\ensuremath{\,\text{Ge\hspace{-.08em}V}}\xspace}
\providecommand{\gev}{\GeV}
\providecommand{\TeV}{\ensuremath{\,\text{Te\hspace{-.08em}V}}\xspace}
\providecommand{\PeV}{\ensuremath{\,\text{Pe\hspace{-.08em}V}}\xspace}
\providecommand{\keVc}{\ensuremath{{\,\text{ke\hspace{-.08em}V\hspace{-0.16em}/\hspace{-0.08em}}c}}\xspace}
\providecommand{\MeVc}{\ensuremath{{\,\text{Me\hspace{-.08em}V\hspace{-0.16em}/\hspace{-0.08em}}c}}\xspace}
\providecommand{\GeVc}{\ensuremath{{\,\text{Ge\hspace{-.08em}V\hspace{-0.16em}/\hspace{-0.08em}}c}}\xspace}
\providecommand{\TeVc}{\ensuremath{{\,\text{Te\hspace{-.08em}V\hspace{-0.16em}/\hspace{-0.08em}}c}}\xspace}
\providecommand{\keVcc}{\ensuremath{{\,\text{ke\hspace{-.08em}V\hspace{-0.16em}/\hspace{-0.08em}}c^\text{2}}}\xspace}
\providecommand{\MeVcc}{\ensuremath{{\,\text{Me\hspace{-.08em}V\hspace{-0.16em}/\hspace{-0.08em}}c^\text{2}}}\xspace}
\providecommand{\GeVcc}{\ensuremath{{\,\text{Ge\hspace{-.08em}V\hspace{-0.16em}/\hspace{-0.08em}}c^\text{2}}}\xspace}
\providecommand{\TeVcc}{\ensuremath{{\,\text{Te\hspace{-.08em}V\hspace{-0.16em}/\hspace{-0.08em}}c^\text{2}}}\xspace}

\providecommand{\pbinv} {\mbox{\ensuremath{\,\text{pb}^\text{$-$1}}}\xspace}
\providecommand{\fbinv} {\mbox{\ensuremath{\,\text{fb}^\text{$-$1}}}\xspace}
\providecommand{\nbinv} {\mbox{\ensuremath{\,\text{nb}^\text{$-$1}}}\xspace}
\providecommand{\percms}{\ensuremath{\,\text{cm}^\text{$-$2}\,\text{s}^\text{$-$1}}\xspace}
\providecommand{\lumi}{\ensuremath{\mathcal{L}}\xspace}
\providecommand{\Lumi}{\ensuremath{\mathcal{L}}\xspace}
%
\providecommand{\LvLow}  {\ensuremath{\mathcal{L}=\text{10}^\text{32}\,\text{cm}^\text{$-$2}\,\text{s}^\text{$-$1}}\xspace}
\providecommand{\LLow}   {\ensuremath{\mathcal{L}=\text{10}^\text{33}\,\text{cm}^\text{$-$2}\,\text{s}^\text{$-$1}}\xspace}
\providecommand{\lowlumi}{\ensuremath{\mathcal{L}=\text{2}\times \text{10}^\text{33}\,\text{cm}^\text{$-$2}\,\text{s}^\text{$-$1}}\xspace}
\providecommand{\LMed}   {\ensuremath{\mathcal{L}=\text{2}\times \text{10}^\text{33}\,\text{cm}^\text{$-$2}\,\text{s}^\text{$-$1}}\xspace}
\providecommand{\LHigh}  {\ensuremath{\mathcal{L}=\text{10}^\text{34}\,\text{cm}^\text{$-$2}\,\text{s}^\text{$-$1}}\xspace}
\providecommand{\hilumi} {\ensuremath{\mathcal{L}=\text{10}^\text{34}\,\text{cm}^\text{$-$2}\,\text{s}^\text{$-$1}}\xspace}


\providecommand{\PT}{\ensuremath{p_{\mathrm{T}}}\xspace}
\providecommand{\pt}{\ensuremath{p_{\mathrm{T}}}\xspace}
\providecommand{\ET}{\ensuremath{E_{\mathrm{T}}}\xspace}
\providecommand{\HT}{\ensuremath{H_{\mathrm{T}}}\xspace}
\providecommand{\et}{\ensuremath{E_{\mathrm{T}}}\xspace}
\providecommand{\Em}{\ensuremath{E\hspace{-0.6em}/}\xspace}
\providecommand{\Pm}{\ensuremath{p\hspace{-0.5em}/}\xspace}
\providecommand{\PTm}{\ensuremath{{p}_\mathrm{T}\hspace{-1.02em}/}\xspace}
\providecommand{\PTslash}{\ensuremath{{p}_\mathrm{T}\hspace{-1.02em}/}\xspace}
\providecommand{\ETm}{\ensuremath{E_{\mathrm{T}}^{\text{miss}}}\xspace}
\providecommand{\MET}{\ETm}
\providecommand{\ETmiss}{\ETm}
\providecommand{\ETslash}{\ensuremath{E_{\mathrm{T}}\hspace{-1.1em}/}\xspace}
\providecommand{\VEtmiss}{\ensuremath{{\vec E}_{\mathrm{T}}^{\text{miss}}}\xspace}

\providecommand{\dd}[2]{\ensuremath{\frac{\mathrm{d} #1}{\mathrm{d} #2}}}
\providecommand{\ddinline}[2]{\ensuremath{\mathrm{d} #1/\mathrm{d} #2}}

\ifthenelse{\boolean{cms@italic}}{\newcommand{\cmsSymbolFace}{\relax}}{\newcommand{\cmsSymbolFace}{\mathrm}}

\providecommand{\zp}{\ensuremath{\cmsSymbolFace{Z}^\prime}\xspace}
\providecommand{\JPsi}{\ensuremath{\cmsSymbolFace{J}\hspace{-.08em}/\hspace{-.14em}\psi}\xspace}
\providecommand{\Z}{\ensuremath{\cmsSymbolFace{Z}}\xspace}
\providecommand{\ttbar}{\ensuremath{\cmsSymbolFace{t}\overline{\cmsSymbolFace{t}}}\xspace}

\newcommand{\cPgn}{\ensuremath{\nu}}
\newcommand{\cPJgy}{\JPsi}
\newcommand{\cPZ}{\Z}
\newcommand{\cPZpr}{\zp}


\providecommand{\AFB}{\ensuremath{A_\text{FB}}\xspace}
\providecommand{\wangle}{\ensuremath{\sin^{2}\theta_{\text{eff}}^\text{lept}(M^2_\Z)}\xspace}
\providecommand{\stat}{\ensuremath{\,\text{(stat.)}}\xspace}
\providecommand{\syst}{\ensuremath{\,\text{(syst.)}}\xspace}
\providecommand{\kt}{\ensuremath{k_{\mathrm{T}}}\xspace}

\providecommand{\BC}{\ensuremath{\mathrm{B_{c}}}\xspace}
\providecommand{\bbarc}{\ensuremath{\mathrm{\overline{b}c}}\xspace}
\providecommand{\bbbar}{\ensuremath{\mathrm{b\overline{b}}}\xspace}
\providecommand{\ccbar}{\ensuremath{\mathrm{c\overline{c}}}\xspace}
\providecommand{\bspsiphi}{\ensuremath{\mathrm{B_s} \to \JPsi\, \phi}\xspace}
\providecommand{\EE}{\ensuremath{\mathrm{e^+e^-}}\xspace}
\providecommand{\MM}{\ensuremath{\mu^+\mu^-}\xspace}
\providecommand{\TT}{\ensuremath{\tau^+\tau^-}\xspace}

\providecommand{\HGG}{\ensuremath{\mathrm{H}\to\gamma\gamma}}
\providecommand{\GAMJET}{\ensuremath{\gamma + \text{jet}}}
\providecommand{\PPTOJETS}{\ensuremath{\mathrm{pp}\to\text{jets}}}
\providecommand{\PPTOGG}{\ensuremath{\mathrm{pp}\to\gamma\gamma}}
\providecommand{\PPTOGAMJET}{\ensuremath{\mathrm{pp}\to\gamma + \mathrm{jet}}}
\providecommand{\MH}{\ensuremath{M_{\mathrm{H}}}}
\providecommand{\RNINE}{\ensuremath{R_\mathrm{9}}}
\providecommand{\DR}{\ensuremath{\Delta R}}

%

\providecommand{\ga}{\ensuremath{\gtrsim}}
\providecommand{\la}{\ensuremath{\lesssim}}
\providecommand{\swsq}{\ensuremath{\sin^2\theta_\cmsSymbolFace{W}}\xspace}
\providecommand{\cwsq}{\ensuremath{\cos^2\theta_\cmsSymbolFace{W}}\xspace}
\providecommand{\tanb}{\ensuremath{\tan\beta}\xspace}
\providecommand{\tanbsq}{\ensuremath{\tan^{2}\beta}\xspace}
\providecommand{\sidb}{\ensuremath{\sin 2\beta}\xspace}
\providecommand{\alpS}{\ensuremath{\alpha_S}\xspace}
\providecommand{\alpt}{\ensuremath{\tilde{\alpha}}\xspace}

\providecommand{\QL}{\ensuremath{\cmsSymbolFace{Q}_\cmsSymbolFace{L}}\xspace}
\providecommand{\sQ}{\ensuremath{\tilde{\cmsSymbolFace{Q}}}\xspace}
\providecommand{\sQL}{\ensuremath{\tilde{\cmsSymbolFace{Q}}_\cmsSymbolFace{L}}\xspace}
\providecommand{\ULC}{\ensuremath{\cmsSymbolFace{U}_\cmsSymbolFace{L}^\cmsSymbolFace{C}}\xspace}
\providecommand{\sUC}{\ensuremath{\tilde{\cmsSymbolFace{U}}^\cmsSymbolFace{C}}\xspace}
\providecommand{\sULC}{\ensuremath{\tilde{\cmsSymbolFace{U}}_\cmsSymbolFace{L}^\cmsSymbolFace{C}}\xspace}
\providecommand{\DLC}{\ensuremath{\cmsSymbolFace{D}_\cmsSymbolFace{L}^\cmsSymbolFace{C}}\xspace}
\providecommand{\sDC}{\ensuremath{\tilde{\cmsSymbolFace{D}}^\cmsSymbolFace{C}}\xspace}
\providecommand{\sDLC}{\ensuremath{\tilde{\cmsSymbolFace{D}}_\cmsSymbolFace{L}^\cmsSymbolFace{C}}\xspace}
\providecommand{\LL}{\ensuremath{\cmsSymbolFace{L}_\cmsSymbolFace{L}}\xspace}
\providecommand{\sL}{\ensuremath{\tilde{\cmsSymbolFace{L}}}\xspace}
\providecommand{\sLL}{\ensuremath{\tilde{\cmsSymbolFace{L}}_\cmsSymbolFace{L}}\xspace}
\providecommand{\ELC}{\ensuremath{\cmsSymbolFace{E}_\cmsSymbolFace{L}^\cmsSymbolFace{C}}\xspace}
\providecommand{\sEC}{\ensuremath{\tilde{\cmsSymbolFace{E}}^\cmsSymbolFace{C}}\xspace}
\providecommand{\sELC}{\ensuremath{\tilde{\cmsSymbolFace{E}}_\cmsSymbolFace{L}^\cmsSymbolFace{C}}\xspace}
\providecommand{\sEL}{\ensuremath{\tilde{\cmsSymbolFace{E}}_\cmsSymbolFace{L}}\xspace}
\providecommand{\sER}{\ensuremath{\tilde{\cmsSymbolFace{E}}_\cmsSymbolFace{R}}\xspace}
\providecommand{\sFer}{\ensuremath{\tilde{\cmsSymbolFace{f}}}\xspace}
\providecommand{\sQua}{\ensuremath{\tilde{\cmsSymbolFace{q}}}\xspace}
\providecommand{\sUp}{\ensuremath{\tilde{\cmsSymbolFace{u}}}\xspace}
\providecommand{\suL}{\ensuremath{\tilde{\cmsSymbolFace{u}}_\cmsSymbolFace{L}}\xspace}
\providecommand{\suR}{\ensuremath{\tilde{\cmsSymbolFace{u}}_\cmsSymbolFace{R}}\xspace}
\providecommand{\sDw}{\ensuremath{\tilde{\cmsSymbolFace{d}}}\xspace}
\providecommand{\sdL}{\ensuremath{\tilde{\cmsSymbolFace{d}}_\cmsSymbolFace{L}}\xspace}
\providecommand{\sdR}{\ensuremath{\tilde{\cmsSymbolFace{d}}_\cmsSymbolFace{R}}\xspace}
\providecommand{\sTop}{\ensuremath{\tilde{\cmsSymbolFace{t}}}\xspace}
\providecommand{\stL}{\ensuremath{\tilde{\cmsSymbolFace{t}}_\cmsSymbolFace{L}}\xspace}
\providecommand{\stR}{\ensuremath{\tilde{\cmsSymbolFace{t}}_\cmsSymbolFace{R}}\xspace}
\providecommand{\stone}{\ensuremath{\tilde{\cmsSymbolFace{t}}_1}\xspace}
\providecommand{\sttwo}{\ensuremath{\tilde{\cmsSymbolFace{t}}_2}\xspace}
\providecommand{\sBot}{\ensuremath{\tilde{\cmsSymbolFace{b}}}\xspace}
\providecommand{\sbL}{\ensuremath{\tilde{\cmsSymbolFace{b}}_\cmsSymbolFace{L}}\xspace}
\providecommand{\sbR}{\ensuremath{\tilde{\cmsSymbolFace{b}}_\cmsSymbolFace{R}}\xspace}
\providecommand{\sbone}{\ensuremath{\tilde{\cmsSymbolFace{b}}_1}\xspace}
\providecommand{\sbtwo}{\ensuremath{\tilde{\cmsSymbolFace{b}}_2}\xspace}
\providecommand{\sLep}{\ensuremath{\tilde{\cmsSymbolFace{l}}}\xspace}
\providecommand{\sLepC}{\ensuremath{\tilde{\cmsSymbolFace{l}}^\cmsSymbolFace{C}}\xspace}
\providecommand{\sEl}{\ensuremath{\tilde{\cmsSymbolFace{e}}}\xspace}
\providecommand{\sElC}{\ensuremath{\tilde{\cmsSymbolFace{e}}^\cmsSymbolFace{C}}\xspace}
\providecommand{\seL}{\ensuremath{\tilde{\cmsSymbolFace{e}}_\cmsSymbolFace{L}}\xspace}
\providecommand{\seR}{\ensuremath{\tilde{\cmsSymbolFace{e}}_\cmsSymbolFace{R}}\xspace}
\providecommand{\snL}{\ensuremath{\tilde{\nu}_L}\xspace}
\providecommand{\sMu}{\ensuremath{\tilde{\mu}}\xspace}
\providecommand{\sNu}{\ensuremath{\tilde{\nu}}\xspace}
\providecommand{\sTau}{\ensuremath{\tilde{\tau}}\xspace}
\providecommand{\Glu}{\ensuremath{\cmsSymbolFace{g}}\xspace}
\providecommand{\sGlu}{\ensuremath{\tilde{\cmsSymbolFace{g}}}\xspace}
\providecommand{\Wpm}{\ensuremath{\cmsSymbolFace{W}^{\pm}}\xspace}
\providecommand{\sWpm}{\ensuremath{\tilde{\cmsSymbolFace{W}}^{\pm}}\xspace}
\providecommand{\Wz}{\ensuremath{\cmsSymbolFace{W}^{0}}\xspace}
\providecommand{\sWz}{\ensuremath{\tilde{\cmsSymbolFace{W}}^{0}}\xspace}
\providecommand{\sWino}{\ensuremath{\tilde{\cmsSymbolFace{W}}}\xspace}
\providecommand{\Bz}{\ensuremath{\cmsSymbolFace{B}^{0}}\xspace}
\providecommand{\sBz}{\ensuremath{\tilde{\cmsSymbolFace{B}}^{0}}\xspace}
\providecommand{\sBino}{\ensuremath{\tilde{\cmsSymbolFace{B}}}\xspace}
\providecommand{\Zz}{\ensuremath{\cmsSymbolFace{Z}^{0}}\xspace}
\providecommand{\sZino}{\ensuremath{\tilde{\cmsSymbolFace{Z}}^{0}}\xspace}
\providecommand{\sGam}{\ensuremath{\tilde{\gamma}}\xspace}
\providecommand{\chiz}{\ensuremath{\tilde{\chi}^{0}}\xspace}
\providecommand{\chip}{\ensuremath{\tilde{\chi}^{+}}\xspace}
\providecommand{\chim}{\ensuremath{\tilde{\chi}^{-}}\xspace}
\providecommand{\chipm}{\ensuremath{\tilde{\chi}^{\pm}}\xspace}
\providecommand{\Hone}{\ensuremath{\cmsSymbolFace{H}_\cmsSymbolFace{d}}\xspace}
\providecommand{\sHone}{\ensuremath{\tilde{\cmsSymbolFace{H}}_\cmsSymbolFace{d}}\xspace}
\providecommand{\Htwo}{\ensuremath{\cmsSymbolFace{H}_\cmsSymbolFace{u}}\xspace}
\providecommand{\sHtwo}{\ensuremath{\tilde{\cmsSymbolFace{H}}_\cmsSymbolFace{u}}\xspace}
\providecommand{\sHig}{\ensuremath{\tilde{\cmsSymbolFace{H}}}\xspace}
\providecommand{\sHa}{\ensuremath{\tilde{\cmsSymbolFace{H}}_\cmsSymbolFace{a}}\xspace}
\providecommand{\sHb}{\ensuremath{\tilde{\cmsSymbolFace{H}}_\cmsSymbolFace{b}}\xspace}
\providecommand{\sHpm}{\ensuremath{\tilde{\cmsSymbolFace{H}}^{\pm}}\xspace}
\providecommand{\hz}{\ensuremath{\cmsSymbolFace{h}^{0}}\xspace}
\providecommand{\Hz}{\ensuremath{\cmsSymbolFace{H}^{0}}\xspace}
\providecommand{\Az}{\ensuremath{\cmsSymbolFace{A}^{0}}\xspace}
\providecommand{\Hpm}{\ensuremath{\cmsSymbolFace{H}^{\pm}}\xspace}
\providecommand{\sGra}{\ensuremath{\tilde{\cmsSymbolFace{G}}}\xspace}
\providecommand{\mtil}{\ensuremath{\tilde{m}}\xspace}
\providecommand{\rpv}{\ensuremath{\rlap{\kern.2em/}R}\xspace}
\providecommand{\LLE}{\ensuremath{LL\bar{E}}\xspace}
\providecommand{\LQD}{\ensuremath{LQ\bar{D}}\xspace}
\providecommand{\UDD}{\ensuremath{\overline{UDD}}\xspace}
\providecommand{\Lam}{\ensuremath{\lambda}\xspace}
\providecommand{\Lamp}{\ensuremath{\lambda'}\xspace}
\providecommand{\Lampp}{\ensuremath{\lambda''}\xspace}
\providecommand{\spinbd}[2]{\ensuremath{\bar{#1}_{\dot{#2}}}\xspace}

\providecommand{\MD}{\ensuremath{{M_\mathrm{D}}}\xspace}
\providecommand{\Mpl}{\ensuremath{{M_\mathrm{Pl}}}\xspace}
\providecommand{\Rinv} {\ensuremath{{R}^{-1}}\xspace} 
\def\ERROR#1#2{ \ensuremath{ \pm #1\, (\textrm{#2}) }  }
\def\RESA#1#2#3{ \ensuremath{ #1 \ERROR{#2}{#3} } }
\def\RESB#1#2#3#4#5{ \ensuremath{ \RESA{#1}{#2}{#3} \ERROR{#4}{#5} } }
\def\RESC#1#2#3#4#5#6#7{ \ensuremath{ \RESB{#1}{#2}{#3}{#4}{#5} \ERROR{#6}{#7} } }
\def\RESD#1#2{ \ensuremath{ #1 \pm #2 } }
\def\EFF#1#2{ \ensuremath{ ( \RESA{#1}{#2}{stat.} )\% } }
\def\EFFA#1#2{ \ensuremath{ ( {#1} \pm {#2} )\% } }
\def\EFFB#1#2#3{ \ensuremath{ ( \RESA{#1}{#2}{stat.} \ERROR{#3}{syst.} )\% } }
\def\SIGBR#1#2{  \ensuremath{ \sigma \left( \pp \to #1 X \right) \times {\cal{B}} \left( #1 \to #2 \right) } }
\def\SIGBRSHORT#1{\ensuremath{ [\, \sigma\times{\cal{B}}\, ](#1) }}
\def\SIGSHORT#1{\ensuremath{ \sigma(#1) }}
\def\RESE#1#2#3#4{ \ensuremath{ \RESC{#1}{#2}{stat.}{#3}{syst.}{#4}{lumi.} }}
\def\RESF#1#2#3{ \ensuremath{ \RESB{#1}{#2}{stat.}{#3}{syst.} }}
\def\RATWZ#1#2{ \ensuremath{ {
 \frac{ \sigma(\pp\rightarrow \PW X)\times {\cal{B}}(\PW\rightarrow #1)  }
      { \sigma(\pp\rightarrow \PZ X)\times {\cal{B}}(\PZ\rightarrow #2)  }   }  } }
\def\RESRATWZ#1#2#3#4#5{ \ensuremath{ \RATWZ{#1}{#2} &=& 
                                   #3 \ERROR{#4}{stat.} \ERROR{#5}{syst.} } }
\def\RATWW#1#2{ \ensuremath{ {
 \frac{ \sigma(\pp\rightarrow \PWp X)\times {\cal{B}}(\PWp\rightarrow #1)  }
      { \sigma(\pp\rightarrow \PWm X)\times {\cal{B}}(\PWm\rightarrow #2)  }   }  } }
\def\RESRATWW#1#2#3#4#5{ \ensuremath{ \RATWW{#1}{#2} &=& 
                                   #3 \ERROR{#4}{stat.} \ERROR{#5}{syst.} } }
\def\THEORYRATIO#1#2{\ensuremath{ \RESD{#1}{#2} }}
\def\EPS#1{ \ensuremath{ \epsilon_{\textrm{#1}} } }
\def\EPSTNPALL#1{ \ensuremath{ \EPS{TNP-WP{#1}-ALL}  }  }
\def\EPSTNPREC{ \ensuremath{ \EPS{TNP-REC} }  }
\def\EPSTNPTRG{ \ensuremath{ \EPS{TNP-TRG} }  }
\def\EPSTNPWP#1{ \ensuremath{ \EPS{TNP-WP{#1} } }  }
\def\EPSTNPTRGWP#1{ \ensuremath{ \EPS{TNP-TRG{#1}} }  } 
\def\XVL#1#2{ \ensuremath{#1_{#2}} }
\def\NSIG#1{ \ensuremath{\XVL{N}{#1} } }
\def\EPSB#1{ \ensuremath{\XVL{\varepsilon}{#1} } }
\def\RHO#1{ \ensuremath{\XVL{\rho}{#1} } }
\def\AGEN#1{ \ensuremath{\XVL{A}{#1} } }
\def\APRIM#1{ \ensuremath{\XVL{ {F} }{#1} } }
\def\RPM{\ensuremath{R_{\scriptscriptstyle{+/-}} }}
\def\RWZ{\ensuremath{R_{\scriptscriptstyle{\PW/\PZ}} }}
\def\LUMI{ \ensuremath{ ( 2.88 \pm 0.32 )~\pbinv  }}
\def\WPWIEFFRECO{   \ensuremath{\EFFA{99.7}{1.0}} } 
\def\WPWIEFFID{     \ensuremath{\EFFA{76.3}{1.9}} } 
\def\WPWIEFFHLT{    \ensuremath{\EFFA{98.9}{1.3}} } 
\def\WPWIEFF{       \ensuremath{\EFFA{75.3}{2.3}} }  
\def\WPWIEFFMC{     \ensuremath{ 79.98\% } } 
\def\WPWITNPR{      \ensuremath{\RESD{0.941}{0.028}} }  
\def\WPWIEBEFFRECO{   \ensuremath{\EFFA{98.6}{0.5}} }  
\def\WPWIEBMCRECO{   \ensuremath{ 98.50\% } }          
\def\WPWIEBRRECO{   \ensuremath{\RESD{1.001}{0.005}} } 

\def\WPWIEBEFFID{   \ensuremath{\EFFA{79.1}{1.8}} }  
\def\WPWIEBMCID{   \ensuremath{ 85.5\% } } 
\def\WPWIEBRID{   \ensuremath{\RESD{0.925}{0.021}} } 

\def\WPZIEBEFFID{   \ensuremath{\EFFA{93.9}{1.5}} } 
\def\WPZIEBMCID{   \ensuremath{ 96.4\% } }  
\def\WPZIEBRID{   \ensuremath{\RESD{0.974}{0.016}} }  

\def\WPWIEBEFFHLT{  \ensuremath{\EFFA{98.9}{0.3}}    } 
\def\WPWIEBMCHLT{   \ensuremath{ 99.70\% }           } 
\def\WPWIEBRHLT{    \ensuremath{\RESD{0.992}{0.003}} }  

\def\WPZIEBEFFHLT{  \ensuremath{\EFFA{98.7}{0.2}}    } 
\def\WPZIEBMCHLT{   \ensuremath{ 99.4\% }           } 
\def\WPZIEBRHLT{    \ensuremath{\RESD{0.992}{0.002}} }  

\def\WPWIEBEFF{    \ensuremath{\EFFA{77.1}{1.8}}     } 
\def\WPWIEBMC{     \ensuremath{ 83.9\% }            } 
\def\WPWIEBR{      \ensuremath{\RESD{0.919}{0.022} } } 

\def\WPZIEBEFF{     \ensuremath{\EFFA{91.3}{1.5}}    } 
\def\WPZIEBMC{     \ensuremath{ 94.4\% }            } 
\def\WPZIEBR{      \ensuremath{\RESD{0.967}{0.016} } } 

\def\WPWIEBSELEFF{       \ensuremath{\EFFA{78.3}{2.9}} }  
\def\WPWIEEEFFRECO{   \ensuremath{\EFFA{96.2}{0.8}} } 
\def\WPWIEEMCRECO{   \ensuremath{ 96.3\% } } 
\def\WPWIEERRECO{   \ensuremath{\RESD{0.999}{0.009}} }  

\def\WPWIEEEFFID{   \ensuremath{\EFFA{69.2}{2.0}} }  
\def\WPWIEEMCID{   \ensuremath{ 74.9\% } } 
\def\WPWIEERID{   \ensuremath{\RESD{0.924}{0.027}} }  

\def\WPWIEEEFFHLT{  \ensuremath{\EFFA{99.2}{0.5}}    } 
\def\WPWIEEMCHLT{   \ensuremath{ 98.80\% }           } 
\def\WPWIEERHLT{    \ensuremath{\RESD{1.003}{0.005}} } 

\def\WPWIEEEFF{     \ensuremath{\EFFA{66.0}{2.0}} }  
\def\WPWIEEMC{     \ensuremath{ 71.3\% } } 
\def\WPWIEER{      \ensuremath{\RESD{0.926}{0.028} } } 

\def\WPZIEEEFFID{   \ensuremath{\EFFA{90.3}{1.9}} } 
\def\WPZIEEMCID{   \ensuremath{ 93.9\% } } 
\def\WPZIEERID{   \ensuremath{\RESD{0.962}{0.020}} }  

\def\WPZIEEEFF{     \ensuremath{\EFFA{86.1}{1.9}} }  
\def\WPZIEEMC{     \ensuremath{ 88.3\% } } 
\def\WPZIEER{      \ensuremath{\RESD{0.975}{0.022} } } 

\def\WPZIEEEFFHLT{   \ensuremath{\EFFA{99.16}{0.02}} } 
\def\WPZIEEMCHLT{   \ensuremath{ 97.7\% } } 
\def\WPZIEERHLT{   \ensuremath{\RESD{1.015}{0.0003}} } 

\def\WPWIEESELEFF{       \ensuremath{\EFFA{66.8}{2.9}} }  
\def\WEITNPDAT{ \ensuremath{\EFFA{73.0}{2.9}} } 
\def\WEPTNPDAT{ \ensuremath{\EFFA{72.5}{3.7}} } 
\def\WEMTNPDAT{ \ensuremath{\EFFA{73.7}{3.7}} } 
\def\WEITNPMC{ \ensuremath{ \EFFA{79.20 }{0.05}} } 
\def\WEPTNPMC{ \ensuremath{ \EFFA{79.03 }{0.07}} } 
\def\WEMTNPMC{ \ensuremath{ \EFFA{79.48 }{0.07}} } 
\def\WEITNPR{ \ensuremath{\RESD{0.921}{0.036} } } 
\def\WEPTNPR{ \ensuremath{\RESD{0.917}{0.046} } } 
\def\WEMTNPR{ \ensuremath{\RESD{0.927}{0.047} } } 
\def\WEIEFFMC{ \ensuremath{\RESD{78.22\%} {0.04\%}} } 
\def\WEPEFFMC{ \ensuremath{\RESD{77.86\%} {0.05\%}} } 
\def\WEMEFFMC{ \ensuremath{\RESD{78.79\%} {0.06\%}} } 

\def\WEIEBEFFMC{ \ensuremath{\RESD{83.05\%} {0.04\%}} }  
\def\WEPEBEFFMC{ \ensuremath{\RESD{82.85\%} {0.05\%}} } 
\def\WEMEBEFFMC{ \ensuremath{\RESD{83.37\%} {0.06\%}} } 

\def\WEIEEEFFMC{ \ensuremath{\RESD{70.05\%} {0.04\%}} } 
\def\WEPEEEFFMC{ \ensuremath{\RESD{69.93\%} {0.06\%}} } 
\def\WEMEEEFFMC{ \ensuremath{\RESD{70.27\%} {0.07\%}} } 

\def\WEIEFF{ \ensuremath{ \EFFA{72.1}{2.8} } } 
\def\WEPEFF{ \ensuremath{ \EFFA{71.4}{3.6} } } 
\def\WEMEFF{ \ensuremath{ \EFFA{73.0}{3.7} } } 
\def\WEISAMPLE{  \ensuremath{28\,601}   }  
\def\WEPSAMPLE{  \ensuremath{15\,859}   }  
\def\WEMSAMPLE{  \ensuremath{12\,742}   }  
\def\WEIAGEN{   \ensuremath{\RESD{0.601}{0.005}} } 
\def\WEPAGEN{   \ensuremath{\RESD{0.622}{0.006}} } 
\def\WEMAGEN{   \ensuremath{\RESD{0.571}{0.009}} } 
\def\WEIAPRIM{   \ensuremath{\RESD{0.446}{0.006}} } 
\def\WEPAPRIM{   \ensuremath{\RESD{0.459}{0.007}} } 
\def\WEMAPRIM{   \ensuremath{\RESD{0.428}{0.008}} } 

\def\WEIACC{   \ensuremath{ \RESD{0.5707 }{ 0.0005 }} } 
\def\WEPACC{   \ensuremath{ \RESD{0.5895 }{ 0.0006 }} } 
\def\WEMACC{   \ensuremath{ \RESD{0.5431 }{ 0.0008 }} } 

\def\WEIEBACC{ \ensuremath{ \RESD{0.3583 }{ 0.0005 }} }  
\def\WEPEBACC{ \ensuremath{ \RESD{0.3618 }{ 0.0006 }} }  
\def\WEMEBACC{ \ensuremath{ \RESD{0.3532 }{ 0.0007 }} }  

\def\WEIEEACC{ \ensuremath{ \RESD{0.2124 }{ 0.0004 }} } 
\def\WEPEEACC{ \ensuremath{ \RESD{0.2277 }{ 0.0005 }} } 
\def\WEMEEACC{ \ensuremath{ \RESD{0.1899 }{ 0.0006 }} } 
\def\WEIYIELD{ \ensuremath{\RESD{11\,895}{115} } } 
\def\WEPYIELD{ \ensuremath{\RESD{ 7\,193}{ 89} } } 
\def\WEMYIELD{ \ensuremath{\RESD{ 4\,728}{ 73} } } 
\def\WEIKSP{ \ensuremath{ 0.00 } } 
\def\WEPKSP{ \ensuremath{ 0.00 } } 
\def\WEMKSP{ \ensuremath{ 0.24 } } 
\def\WEIKSPCOR{ \ensuremath{ 0.49 } } 
\def\WEPKSPCOR{ \ensuremath{ 0.39 } } 
\def\WEMKSPCOR{ \ensuremath{ 0.53 } } 
\def\ZEESAMPLE{  \ensuremath{673}   } 
\def\ZEESAMPLEN{  \ensuremath{677}   } 

\def\ZEEYIELD{ \ensuremath{ \RESD{674}{26} } } 
\def\ZEEQCDBKG{ \ensuremath{ \RESD{0.4}{0.4}  } } 
\def\ZEEEWKBKG{   \ensuremath{ \RESD{2.36}{0.04}   } } 
\def\ZEEBKG{ \ensuremath{ \RESD{2.8}{0.4}  } } 
\def\ZEEAGEN{     \ensuremath{ \RESD{0.479}{0.005} } }  
\def\ZEEAPRIM{     \ensuremath{ \RESD{0.285}{0.005} } } 

\def\ZEEACC{     \ensuremath{ \RESD{0.4345}{0.0005} } } 
\def\ZEEBBACC{   \ensuremath{ \RESD{0.2257}{0.0004} } } 
\def\ZEEBEACC{   \ensuremath{ \RESD{0.1612}{0.0003} } } 
\def\ZEEEEACC{   \ensuremath{ \RESD{0.0476}{0.0002} } } 

\def\ZEBEBFRAC{       \ensuremath{52\%} }  
\def\ZEBEEFRAC{       \ensuremath{37\%} }  
\def\ZEEEEFRAC{       \ensuremath{11\%} }  

\def\ZEETNPDAT{ \ensuremath{ \EFFA{55.7}{3.3} } }  
\def\ZEETNPMC{ \ensuremath{ \EFFA{65.04}{0.05} } }  
\def\ZEETNPR{  \ensuremath{ \RESD{0.856}{0.050} } }  
\def\ZEEEFFMC{ \ensuremath{ \EFFA{65.62}{0.07} } }  
\def\ZEEEFF{   \ensuremath{ \EFFA{56.2}{3.3} } }  

\def\EBESCALE{ \ensuremath{ {1.015}\pm{0.002} } } 
\def\EEESCALE{ \ensuremath{ {1.033}\pm{0.005} } }  

\def\EBESMEAR{ \ensuremath{ {0.82}\pm{0.16}~\GeV } } 
\def\EEESMEAR{ \ensuremath{ {0.67}\pm{0.35}~\GeV } }  
\def\WEITNPSYST{ \ensuremath{ 3.9 } }
\def\WEIESCALESYST{ \ensuremath{ 2.0 } }
\def\WEPESCALESYST{ \ensuremath{ 2.2 } }
\def\WEMESCALESYST{ \ensuremath{ 1.8 } }
\def\WERESCALESYST{ \ensuremath{ 0.4 } }
\def\WEIMETSYST{ \ensuremath{ 1.8 } }
\def\WEPMETSYST{ \ensuremath{ 1.6 } }
\def\WEMMETSYST{ \ensuremath{ 1.9 } }
\def\WERMETSYST{ \ensuremath{ 0.4 } }
\def\WEIBKGSYST{ \ensuremath{ 1.3 } }
\def\WEPBKGSYST{ \ensuremath{ 1.1 } }
\def\WEMBKGSYST{ \ensuremath{ 1.5 } }
\def\WERBKGSYST{ \ensuremath{ 0.7 } }
\def\ZEEBKGSYST{ \ensuremath{ 0.1 } }
\def\WPWMISID{ \ensuremath{ 1.2^{+0.4}_{-0.3} } }

\def\ZEEESCALESYST{ \ensuremath{ 0.6 } }
\def\ZEETNPSYST{ \ensuremath{ 5.9 } }

\def\WEIPDFACCSYST{\ensuremath{ 0.8 }}
\def\ZEEPDFACCSYST{\ensuremath{ 1.1 }}
\def\WEITHSYST{\ensuremath{ 1.3 }}
\def\ZEETHSYST{\ensuremath{ 1.3 }}
\def\WEITOTSYST{\ensuremath{ 5.1 }}
\def\ZEETOTSYST{\ensuremath{ 6.2 }}
\newcommand{\THELUMI} {\ensuremath{{2.88\pm 0.32}~\mathrm{pb}^{-1}}}%

\def\WMUIEFFSA{      \ensuremath{\EFFA{96.4}{0.5}} }
\def\WMUIMCEFFSA{    \ensuremath{ 97.2\% }  } 
\def\WMUIRSA{      \ensuremath{\RESD{0.992}{0.005}}  }
\def\WMUIEFFTRK{      \ensuremath{\EFFA{99.1}{0.4}} }
\def\WMUIMCEFFTRK{    \ensuremath{ 99.3\% }  } 
\def\WMUIRTRK{      \ensuremath{\RESD{0.998}{0.003}}  }
\def\WMUIEFFSEL{      \ensuremath{\EFFA{99.7}{0.3}} }
\def\WMUIMCEFFSEL{    \ensuremath{ 99.7\% }  } 
\def\WMUIRSEL{      \ensuremath{\RESD{1.000}{0.003}}  }
\def\WMUIEFFISO{      \ensuremath{\EFFA{98.5}{0.4}} }
\def\WMUIMCEFFISO{    \ensuremath{ 99.1\% }  } 
\def\WMUIRISO{      \ensuremath{\RESD{0.994}{0.004}}  }
\def\WMUIEFFTRG{      \ensuremath{\EFFA{88.3}{0.8}} }
\def\WMUIMCEFFTRG{    \ensuremath{ 93.2\% }  } 
\def\WMUIRTRG{      \ensuremath{\RESD{0.947}{0.009}}  }
\def\WMUIEFF{      \ensuremath{\EFFA{82.8}{1.0}} }
\def\WMUIMCEFF{    \ensuremath{ 88.7\% }  } 

\def\WMIEFFPLS{\ensuremath{ \RESD{0.935}{0.018} }}
\def\WMIEFFMIN{\ensuremath{ \RESD{0.931}{0.019} }}
\def\WMIEFFBAR{\ensuremath{ \RESD{0.955}{0.024} }}
\def\WMIEFFTRA{\ensuremath{ \RESD{0.89}{0.04} }}
\def\WMIEFFEND{\ensuremath{ \RESD{0.92}{0.03} }}
\def\WMIRHOEFF{ \ensuremath{\RESD{0.933}{0.012}} }
\def\WMUIR{        \ensuremath{\WMIRHOEFF}  }
\def\WMIAGEN{   \ensuremath{\RESD{0.525}{0.006}} } 
\def\WMPAGEN{   \ensuremath{\RESD{0.541}{0.006}} } 
\def\WMMAGEN{   \ensuremath{\RESD{0.502}{0.006}} } 
\def\ZMMAGEN{   \ensuremath{\RESD{0.398}{0.005}} }

\def\WMIAPRIM{  \ensuremath{\RESD{0.462}{0.005}} } 
\def\WMPAPRIM{  \ensuremath{\RESD{0.477}{0.005}} } 
\def\WMMAPRIM{  \ensuremath{\RESD{0.441}{0.005}} } 
\def\ZMMBG{ \ensuremath{\RESD{3.5}{0.2}} }
\def\WMISAMPLE{  \ensuremath{18\,571} }
\def\WMPSAMPLE{  \ensuremath{10\,682} }
\def\WMMSAMPLE{  \ensuremath{ 7\,889} }
\def\WMISAMPLEH{ \ensuremath{11\,011} }
\def\WMPSAMPLEH{ \ensuremath{ 6\,495} }
\def\WMMSAMPLEH{ \ensuremath{ 4\,516} }
\def\ZMMSAMPLE{  \ensuremath{913} }
\def\WMIYIELD{ \ensuremath{\RESD{12\,257}{111} }} 
\def\WMPYIELD{ \ensuremath{\RESD{ 7\,445}{ 87} }} 
\def\WMMYIELD{ \ensuremath{\RESD{ 4\,812}{ 69} }} 
\def\ZMMYIELD{ \ensuremath{\RESD{1\,050}{35} }} 
\def\WMIKSPCOR{ \ensuremath{ 0.34 } } 
\def\WMPKSPCOR{ \ensuremath{ 0.19 } } 
\def\WMMKSPCOR{ \ensuremath{ 0.23 } } 
\def\WMIEFFSYST{     \ensuremath{ 1.4 }}
\def\WMIEFFPRET{     \ensuremath{ 0.5 }}
\def\ZMMEFFSYST{     \ensuremath{ 1.2 }}
\def\ZMMEFFPRET{     \ensuremath{ 0.5 }}
\def\WMIEFFSYSTPRET{     \ensuremath{ 1.5 }}
\def\WMISCALESYST{   \ensuremath{ 0.3 }}
\def\ZMMSCALESYST{   \ensuremath{ 0.2 }}
\def\WMIQCDSHAPESYST{\ensuremath{ 2.0 }}

\def\WMIMETSYST{ \ensuremath{ 0.4 } }
\def\WMIBKGSYST{\ensuremath{ ? }}
\def\ZMMFITSYST{\ensuremath{ 1.0 }}
\def\ZMMBKGSYST{\ensuremath{ 0.2 }}

\def\WMIPDFACCSYST{\ensuremath{ 1.1 }}
\def\ZMMPDFACCSYST{\ensuremath{ 1.2 }}
\def\WMITHSYST{    \ensuremath{ 1.4 }}
\def\ZMMTHSYST{    \ensuremath{ 1.6 }}
\def\WMITOTSYST{   \ensuremath{ 3.1 }}
\def\ZMMTOTSYST{   \ensuremath{ 2.3 }}
\def\THEORYSIGBR#1#2{\ensuremath{ \RESD{#1}{#2}~{\mathrm{nb}} }}
\def\THEORYSIGBRWI{\ensuremath{ \THEORYSIGBR{10.44}{0.52} }}
\def\THEORYSIGBRWP{\ensuremath{ \THEORYSIGBR{6.15}{0.29} }}
\def\THEORYSIGBRWM{\ensuremath{ \THEORYSIGBR{4.29}{0.23} }}
\def\THEORYSIGBRZ{ \ensuremath{ \THEORYSIGBR{0.97}{0.04} }}
\def\THEORYRATIOWZ{\ensuremath{ \THEORYRATIO{10.74}{0.04} }}
\def\THEORYRATIOWW{\ensuremath{ \THEORYRATIO{1.43}{0.04} }}
\def\RATCMSTHY#1#2#3{\ensuremath{ #1\pm #2\,{\mathrm{(exp.)}}\pm #3\,{\mathrm{(theo.)}} }}
\def\RATCMSTHYWI{\ensuremath{\RATCMSTHY{0.953}{0.028}{0.048}}}
\def\RATCMSTHYWP{\ensuremath{\RATCMSTHY{0.953}{0.029}{0.045}}}
\def\RATCMSTHYWM{\ensuremath{\RATCMSTHY{0.954}{0.034}{0.051}}}
\def\RATCMSTHYZ {\ensuremath{\RATCMSTHY{0.960}{0.036}{0.040}}}
\def\RATCMSTHYWZ{\ensuremath{\RATCMSTHY{0.990}{0.038}{0.004}}}
\def\RATCMSTHYWW{\ensuremath{\RATCMSTHY{1.002}{0.038}{0.028}}}
\def\WEISIGBRGHM{ \ensuremath{  \RESE
                             {10.04}{0.10}{0.52}{1.10} } }  
\def\WEPSIGBRGHM{ \ensuremath{  \RESE
                             {5.93}{0.07}{0.36}{0.65} } } 
\def\WEMSIGBRGHM{ \ensuremath{  \RESE
                             {4.14}{0.06}{0.25}{0.45} } } 
\def\WMISIGBRGHM{\ensuremath{ \RESE
                             {9.92}{0.09}{0.31}{1.09}} }
\def\WMPSIGBRGHM{\ensuremath{ \RESE
                             {5.84}{0.07}{0.18}{0.64} } }
\def\WMMSIGBRGHM{\ensuremath{ \RESE
                             {4.08}{0.06}{0.15}{0.45} } }
\def\RESRATWZGHME{ \ensuremath{ \RESF
                             {10.47}{0.42}{0.47} } }
\def\RESRATWZGHMM{\ensuremath{ \RESF
                            {10.74}{0.37}{0.33} }}
\def\WLISIGBRGHM{\ensuremath{ \RESE
                             {9.95}{0.07}{0.28}{1.09} } }
\def\WLPSIGBRGHM{\ensuremath{ \RESE
                             {5.86}{0.06}{0.17}{0.64} } }
\def\WLMSIGBRGHM{\ensuremath{ \RESE
                             {4.09}{0.05}{0.14}{0.45} } }
\def\RESRATWZGHML{ \ensuremath{ \RESF
                             {10.64}{0.28}{0.29} }}

\def\ZEESIGBRGHM{ \ensuremath{  \RESE
                             {0.960}{0.037}{0.059}{0.106} } } 
\def\ZMMSIGBRGHM{\ensuremath{ \RESE
                             {0.924}{0.031}{0.022}{0.102} } }
\def\RESRATWWGHME{ \ensuremath{ \RESF
                             {1.434}{0.028}{0.082} }}
\def\RESRATWWGHMM{\ensuremath{ \RESF
                            {1.433}{0.026}{0.054} }}
\def\ZLLSIGBRGHM{\ensuremath{ \RESE 
                             {0.931}{0.026}{0.023}{0.102} } }
\def\RESRATWWGHML{ \ensuremath{ \RESF
                             {1.433}{0.020}{0.050} }}

\def\THGHMSIGBRWI{\ensuremath{ \RESD{10.44}{0.52} }}
\def\THGHMSIGBRWP{\ensuremath{ \RESD{6.15}{0.29} }}
\def\THGHMSIGBRWM{\ensuremath{ \RESD{4.29}{0.23} }}
\def\THGHMSIGBRZ{ \ensuremath{ \RESD{0.972}{0.042} }}
\def\THGHMRATIOWZ{\ensuremath{ \RESD{10.74}{0.04} }}
\def\THGHMRATIOWW{\ensuremath{ \RESD{1.43}{0.04} }}

\def\WEISIGBRXGHM{\ensuremath{ \RESE
                             {6.04}{0.06}{0.31}{0.66} } }
\def\WEPSIGBRXGHM{\ensuremath{ \RESE
                             {3.69}{0.05}{0.22}{0.41} } }
\def\WEMSIGBRXGHM{\ensuremath{ \RESE
                             {2.36}{0.04}{0.14}{0.26} } }
\def\WMISIGBRXGHM{\ensuremath{ \RESE
                             {5.21}{0.05}{0.15}{0.57} } }
\def\WMPSIGBRXGHM{\ensuremath{ \RESE
                             {3.16}{0.04}{0.10}{0.35} } }
\def\WMMSIGBRXGHM{\ensuremath{ \RESE
                             {2.05}{0.03}{0.06}{0.22} } }

\def\ZEESIGBRXGHM{\ensuremath{ \RESE
                             {0.460}{0.018}{0.028}{0.051} } }
\def\ZMMSIGBRXGHM{\ensuremath{ \RESE
                             {0.368}{0.012}{0.007}{0.040} } }
\renewcommand{\PJgy}{\ensuremath{\mathrm{J}\hspace{-.08em}/\hspace{-.14em}\psi}}
\renewcommand{\ttbar}{\ensuremath{\mathrm{t}\overline{\mathrm{t}}}}
\newcommand{\pp}{\ensuremath{{\Pp\Pp}}}%
\newcommand{\PZ}{\ensuremath{{\mathrm{Z}}}}%
\newcommand{\WW}{\ensuremath{\PW\PW}}%
\newcommand{\WZ}{\ensuremath{\PW\PZ}}%
\newcommand{\ZZ}{\ensuremath{\PZ\PZ}}%
\newcommand{\rts}{\ensuremath{\sqrt{s}}}%
\newcommand{\ra}{\ensuremath{\rightarrow}}%

\newcommand{\MN}{\ensuremath{\Pgm\nu}}%
\newcommand{\MpN}{\ensuremath{\Pgmp\nu}}%
\newcommand{\MmN}{\ensuremath{\Pgmm\overline{\nu}}}%
\newcommand{\EN}{\ensuremath{\Pe\nu}}%
\newcommand{\EpN}{\ensuremath{\Pep\nu}}%
\newcommand{\EmN}{\ensuremath{\Pem\overline{\nu}}}%
\newcommand{\TN}{\ensuremath{\Pgt\nu}}%
\newcommand{\TpN}{\ensuremath{\Pgt^+\nu}}%
\newcommand{\TmN}{\ensuremath{\Pgt^-\overline{\nu}}}%
\newcommand{\LN}{\ensuremath{\ell\nu}}%
\newcommand{\LpN}{\ensuremath{\ell^+\nu}}%
\newcommand{\LmN}{\ensuremath{\ell^-\overline{\nu}}}%

\newcommand{\MpMm}{\ensuremath{\mu^+\mu^-}}%
\newcommand{\TpTm}{\ensuremath{\tau^+\tau^-}}%
\newcommand{\LpLm}{\ensuremath{\ell^+\ell^-}}%
\newcommand{\EpEm}{\ensuremath{\Pep\Pem}}%
\newcommand{\MW}{\ensuremath{{m}_\PW}}%
\newcommand{\MZ}{\ensuremath{{m}_\PZ}}%
\newcommand{\MT}{\ensuremath{{M}_{\mathrm{T}}}}%
\newcommand{\MLL}{\ensuremath{{M}_{\LpLm}}}%

\newcommand{\Wmn}{\ensuremath{\PW \ra \MN}}%
\newcommand{\Wpmn}{\ensuremath{\PWp \ra \MpN}}%
\newcommand{\Wmmn}{\ensuremath{\PWm \ra \MmN}}%
\newcommand{\Zmm}{\ensuremath{\PZ \ra \MpMm}}%
\newcommand{\ppWmn}{\pp \ra \PW + X \ra \MN + X}%
\newcommand{\ppWpmn}{\pp \ra \PWp + X \ra \MpN + X}%
\newcommand{\ppWmmn}{\pp \ra \PWm + X \ra \MmN + X}%
\newcommand{\ppZmm}{\pp \ra \PZ + X \ra \MpMm + X}%

\newcommand{\Wen}{\ensuremath{\PW \ra \EN}}%
\newcommand{\Wpen}{\ensuremath{\PWp \ra \EpN}}%
\newcommand{\Wmen}{\ensuremath{\PWm \ra \EmN}}%
\newcommand{\Zee}{\ensuremath{\PZ \ra \EpEm}}%
\newcommand{\ppWen}{\pp \ra \PW + X \ra \EN + X}%
\newcommand{\ppWpen}{\pp \ra \PWp + X \ra \EpN  + X}%
\newcommand{\ppWmen}{\pp \ra \PWm + X \ra \EmN + X}%
\newcommand{\ppZee}{\pp \ra \PZ + X \ra \EpEm + X}%

\newcommand{\Wtn}{\ensuremath{\PW \ra \TN}}%
\newcommand{\Wptn}{\ensuremath{\PWp \ra \TpN}}%
\newcommand{\Wmtn}{\ensuremath{\PWm \ra \TmN}}%
\newcommand{\Ztt}{\ensuremath{\PZ \ra \TpTm}}%

\newcommand{\Wln}{\ensuremath{\PW \ra \LN}}%
\newcommand{\Wpln}{\ensuremath{\PWp \ra \LpN}}%
\newcommand{\Wmln}{\ensuremath{\PWm \ra \LmN}}%
\newcommand{\Zll}{\ensuremath{\PZ \ra \LpLm}}%
\newcommand{\ppZll}{\pp \ra \PZ + X \ra \LpLm + X}%
\newcommand{\ppWln}{\pp \ra \PW + X \ra \LN + X}%
\newcommand{\ppWpln}{\pp \ra \PWp + X \ra \LpN  + X}%
\newcommand{\ppWmln}{\pp \ra \PWm + X \ra \LmN + X}%

\renewcommand{\MET}{\ensuremath{{E\!\!\!/}_{\mathrm{T}}}\xspace}
\newcommand{\gammaZ}{\ensuremath{\PZ/\gamma^{*}}}
\newcommand{\gammaZmm}{\mbox{$ \gammaZ\rightarrow \MpMm$}}
\newcommand{\gammaZee}{\mbox{$ \gammaZ\rightarrow \EpEm$}}
\newcommand{\gammaZtt}{\mbox{$ \gammaZ\rightarrow \TpTm$}}
\newcommand{\gammaZll}{\mbox{$ \gammaZ\rightarrow \LpLm$}}

\newcommand{\mz}{\mbox{$m_{\PZ}$}}
\newcommand{\hta}{\mbox{$\eta$}}
\newcommand{\fh}{\mbox{$\phi$}}
\newcommand{\etot}{\mbox{$\epsilon_{tot}$}}
\newcommand{\eclustering}{\mbox{$ \epsilon_{clustering}$}}
\newcommand{\etracking}{\mbox{$ \epsilon_{tracking}$}}
\newcommand{\egsfele}{\mbox{$ \epsilon_{gsfele}$}}
\newcommand{\epreselection}{\mbox{$ \epsilon_{preselection}$}}
\newcommand{\eisolation}{\mbox{$ \epsilon_{isolation}$}}
\newcommand{\eclassification}{\mbox{$ \epsilon_{classification}$}}
\newcommand{\eelID}{\mbox{$ \epsilon_{elID}$}}
\newcommand{\etrigger}{\mbox{$ \epsilon_{trigger}$}}

\newcommand{\DE}{$\Delta\eta_{in}$}
\newcommand{\DP}{$\Delta\phi_{in}$}
\newcommand{\SEE}{$\sigma_{\eta\eta}$~}
\newcommand{\SEP}{$\sigma_{\eta\phi}$}
\newcommand{\SPP}{$\sigma_{\phi\phi}$}
\newcommand{\SXY}{$\sigma_{XY}$}
\newcommand{\pth}{\hat{p}_{\perp}}
\newcommand{\Lint}{\ensuremath{{\cal L}_{\mathrm{int}}}}
\newcommand{\IECAL}    {I^{\textrm{rel}}_{\scriptscriptstyle{\textrm{ECAL}}}}%
\newcommand{\IHCAL}    {I^{\textrm{rel}}_{\scriptscriptstyle{\textrm{HCAL}}}}%
\newcommand{\ITRK}     {I^{\textrm{rel}}_{\scriptscriptstyle{\textrm{trk}}}}%
\newcommand{\IRelComb} {I^{\textrm{rel}}_{\scriptscriptstyle{\textrm{comb}}}}%
\newcommand{\Nev}    {N_{\mathrm{ev}}}
\newcommand{\Nsig}   {N_{\mathrm{sig}}}
\newcommand{\Nsel}   {N_{\mathrm{sel}}}
\newcommand{\Nbg}    {N_{\mathrm{bg}}}
\newcommand{\rhoeff} {\rho_{\mathrm{eff}}}
\newcommand{\effmc}  {\epsilon_{\mathrm{sim}}}
\newcommand{\effdt}  {\epsilon_{\mathrm{data}}}
\newcommand{\etaSC}  {\eta_{\mathrm{SC}}}
\cmsNoteHeader{EWK-10-002}

\title{Measurements of Inclusive $\mathrm{W}$ and $\mathrm{Z}$ Cross Sections\\ 
in $\mathrm{pp}$ Collisions at $\sqrt{s}=7$~TeV}

\address[cern]{CERN}
\author[cern]{The CMS Collaboration}

\date{\today}
\abstract{
Measurements of inclusive  $\mathrm{W}$ and $\mathrm{Z}$ boson production 
cross sections in $\pp$ collisions 
at $\sqrt{s}=7~\mathrm{TeV}$
are presented,
based on $2.9~\mathrm{pb}^{-1}$ of data 
recorded by 
the CMS detector
at the 
LHC. 
The measurements, performed in the electron and muon decay channels,
are combined to give
$\sigma( \mathrm{pp} \rightarrow \mathrm{W}X ) 
\times {\cal{B}}( \mathrm{W} \rightarrow \ell \nu ) =
9.95\pm0.07\,{\textrm{(stat.)}}\pm 0.28\,{\textrm{(syst.)}}\pm1.09\,{\textrm{(lumi.)}}~\mathrm{nb}$ 
and
$\sigma( \mathrm{pp} \rightarrow \mathrm{Z}X ) 
\times {\cal{B}}( \mathrm{Z} \rightarrow \ell^+ \ell^- ) =
0.931\pm0.026\,{\textrm{(stat.)}}\pm 0.023\,{\textrm{(syst.)}}\pm0.102\,{\textrm{(lumi.)}}~\mathrm{nb}$,
where $\ell$ stands for either $\mathrm{e}$ or $\mu$.
Theoretical predictions, calculated 
at the next-to-next-to-leading order in 
QCD using 
recent
parton distribution functions,
are in agreement with the measured cross sections.
Ratios of cross sections,
which incur an experimental systematic uncertainty of less than $4\%$, 
are also reported.
}
\hypersetup{%
pdfauthor={CMS Collaboration},%
pdftitle={Measurements of Inclusive W and Z Cross Sections in pp Collisions at sqrt(s)=7 TeV},%
pdfsubject={CMS},%
pdfkeywords={CMS, physics}}

\maketitle 
\section{Introduction}
\par
The inclusive production of $\PW$ and $\PZ$ bosons is an important
benchmark process at hadron colliders.
Measurements of
$ \SIGBR{\PW}{\LN}$ and
$ \SIGBR{\PZ}{\LpLm}$, 
where $\ell=\Pe$ or $\Pgm$,
test calculations based on higher-order perturbative QCD and
parton distribution functions (PDF).  Such calculations
are supported by measurements 
at the S$\Pap\Pp$S~\cite{Albajar:1988ka,Alitti:1990gj}
and Tevatron 
\cite{Acosta:2004uq,Affolder:1999jh,Abbott:1999tt}
$\Pp\Pap$ colliders.
We report the extension of
these measurements
to significantly higher energies, namely, with $\pp$
collisions at a center-of-mass energy of $7$~TeV 
provided by the Large Hadron Collider (LHC).
The data were collected from April through August, 2010, by the
Compact Muon Solenoid (CMS) experiment, and correspond to an 
integrated luminosity of $\LUMI$.
Recently, the ATLAS Collaboration published 
measurements of cross sections for inclusive $\PW$ and $\PZ$
productions at the LHC based on approximately 
$0.34~\textrm{pb}^{-1}$~\cite{ATLASWZ}.
In this article, ``$\PZ$ boson production'' 
includes $\gamma^*$ exchange within the mass range 60 to 120~GeV.

High-$\pt$ electrons and muons are used for selecting 
$\Wln$ and $\Zll$
candidate events. 
In addition to a high-$\pt$ lepton, $\PW$ events are characterized 
by significant missing transverse energy ($\MET$) due to the 
escaping neutrino.
The reconstruction of electrons
and muons is detailed in Section~\ref{sec:leptons}, 
along with lepton identification and isolation requirements, and
the $\MET$ reconstruction is described in Section~\ref{sec:MET}.

The measurements of cross sections are based on the formula 
$\sigma \times {\cal B} 
= \NSIG{} / ( \AGEN{}  \times \EPSB{} \times \lumi )$, 
where  $\NSIG{}$ is the number of signal events, 
$\AGEN{}$ is the fiducial and kinematic acceptance, 
$\EPSB{}$ is the selection efficiency for events in the acceptance, 
and  ${\lumi}$ is the integrated luminosity. 
The value of $\AGEN{}$ is affected by PDF and
renormalization scale uncertainties, while the value of  $\EPSB{}$ is
susceptible to errors from triggering and reconstruction.  In
order to control the efficiency uncertainties, we concentrate on
the extraction of corrections to the efficiencies obtained from
the simulation; these correction factors come from efficiency
ratios  $\RHO{}= \EPSB{}/{\EPSB{}}_{\mathrm{sim}}$
derived by measuring $\EPSB{}$ and ${\EPSB{}}_{\mathrm{sim}}$ in the same way
on data and simulations, respectively.  
In effect, we replace the product $\AGEN{}  \times \EPSB{}$
by the product $\APRIM{} \times \RHO{}$, 
where  $\APRIM{}=A \times {\EPSB{}}_{\mathrm{sim}}$
is the fraction of generated events selected in the simulation.
The values for $\RHO{}$ are derived from data, and hence their
uncertainties are experimental; the uncertainties on $\APRIM{}$
derive from the theoretical uncertainties on $\AGEN{}$. 
In order to exploit this distinction between experimental and theoretical
uncertainties, we also report cross section measurements defined within
the restricted acceptance dictated by the detector coverage and
minimum transverse momentum; these values incur essentially no
theoretical uncertainty.

\par
In Section~\ref{sec:efficiencies} we 
determine electron and muon selection efficiency correction factors 
from the data. The selection of events for
the $\PW$ and $\PZ$ samples and the 
extraction of signal event yields 
are outlined 
in Section~\ref{sec:selection}, followed by a discussion of  
systematic uncertainties in Section~\ref{sec:systematics}.
Finally, the results are reported 
and briefly discussed in Section~\ref{sec:results}.

\par
In the following
section, 
a brief description of the CMS
detector is provided. 
\section{The CMS detector}
\label{sec:CMS}
\par
The central feature of the CMS apparatus 
is a superconducting solenoid of 6~m internal diameter, providing 
a magnetic field of $3.8$~T. Within the field volume are a silicon pixel 
and strip tracker, an electromagnetic calorimeter (ECAL) 
and a brass/scintillator hadron calorimeter (HCAL). Muons are detected 
in gas-ionization detectors embedded in the steel return 
yoke. In addition to the barrel and endcap detectors, CMS has 
extensive forward calorimetry.
\par
CMS uses a right-handed coordinate system, with the origin at the 
nominal interaction point, the $x$-axis pointing to the center of 
the LHC ring, the $y$-axis pointing up (perpendicular to the LHC plane), 
and the $z$-axis along the anticlockwise-beam direction. The polar 
angle $\theta$ is measured from the positive $z$-axis and the 
azimuthal angle $\phi$ is measured in radians in the $xy$-plane.
The pseudorapidity is given by $\eta = -\ln(\tan\theta/2)$.
\par
The inner tracker measures charged particle trajectories in the 
pseudorapidity range $|\eta| < 2.5$.   It consists of $1\,440$ silicon 
pixel and 15\,148 silicon strip detector modules.  It provides an 
impact parameter resolution of $\sim 15\mum$ and a transverse 
momentum ($\pt$) resolution of about 1\% for charged particles 
with $\pt \approx 40\GeV$.
\par
The electromagnetic calorimeter consists of nearly $76\,000$ lead tungstate
crystals which provide coverage in pseudorapidity $|\eta| < 1.479$ in a 
cylindrical barrel region (EB) and $1.479 < |\eta| < 3.0$ in two endcap 
regions (EE).  
A preshower detector
consisting of two planes of silicon sensors interleaved with a total of
3~X$_0$ of lead is located in front of the EE.
The ECAL has an ultimate energy resolution of better than $0.5\%$ for
unconverted photons with transverse energies above $100~\GeV$.
The energy resolution is $3\%$ or better for the range of
electron energies relevant for this analysis. 
The hadronic calorimeter is a sampling device with brass
as passive material and scintillator as active material.
The combined calorimeter cells are grouped in projective towers of granularity 
$\Delta \eta \times \Delta \phi = 0.087\times0.087$ at central rapidities 
and $0.175\times0.175$ at forward rapidities.
\par
Muons are detected in the pseudorapidity window $|\eta|< 2.4$, with 
detection planes based on three technologies: drift tubes, cathode strip 
chambers, and resistive plate chambers.  A high-$\pt$ muon originating
from the interaction point produces track segments in typically 
three or four muon stations.  Matching these segments to tracks 
measured in the inner tracker results in a $\pt$ resolution 
between 1 and 2\% for $\pt$ values up to $100$~GeV.
\par
The first level (L1) of the CMS trigger system, composed of custom 
hardware processors, is designed to select the most interesting events
in less than $1\mus$ 
using information from the calorimeters 
and muon detectors. The High Level Trigger (HLT) processor farm further 
decreases the event rate 
to a few hundred hertz, 
before data storage.
\par
A more detailed description of CMS can be found elsewhere~\cite{JINST}.
\section{Lepton Reconstruction and Identification}
\label{sec:leptons}
\par
Events in which hadronic jets mimic an electron or a muon can contaminate
the $\PW$ and $\PZ$ samples.   
Such fake leptons, 
as well as real leptons arising from decays of heavy-flavour hadrons
or decays in flight of light mesons
within jets, are suppressed by imposing limits on additional 
energy recorded near the projected impact point of the candidate
lepton in the calorimeters, as well as on the energy of charged particles
reconstructed in the inner tracker near the direction of the
candidate lepton.  
We define 
isolation
variables for the three subsystems:  
$\IECAL = \sum \et(\textrm{ECAL})/\pt^{\ell}$, 
$\IHCAL = \sum \et(\textrm{HCAL})/\pt^{\ell}$ and 
$\ITRK  = \sum \pt(\textrm{tracks})/\pt^{\ell}$,
where  $\pt^{\ell}$ is the transverse momentum of the lepton candidate.
The scalar sums of transverse energy ($\et$) 
and transverse momentum ($\pt$) 
are performed for objects falling within a cone
$\Delta R = \sqrt{(\Delta\eta)^2+(\Delta\phi)^2} < 0.3$ around
the lepton candidate,
the energy deposits and the track associated with the lepton candidate 
being excluded from the sums.  We also define a combined isolation variable,
$\IRelComb =  \IECAL+\IHCAL+\ITRK $.
\subsection{Electrons}
\par
Events with high-$\et$ electrons are selected online when they pass a L1 trigger 
filter that requires a coarse-granularity region of the ECAL to have 
$\et > 5$~GeV. They subsequently must pass an HLT~\cite{HLT}
filter that requires an ECAL cluster with $\et > 15$~GeV, using
the full granularity of the ECAL and $\et$ measurements 
corrected using 
offline calibration~\cite{CMS-PAS-EGM-10-003}.
\par
Electrons are identified offline as clusters of ECAL energy deposits 
matched to tracks from the silicon tracker. The ECAL clusters are designed
to collect the largest fraction of the energy of the original electron,
including energy radiated along its trajectory.  They must
fall in the ECAL fiducial volume
of
$|\eta| < 1.44$ for EB
clusters or $1.57 < |\eta| < 2.5$ for EE clusters. 
The transition region 
from $1.44 < |\eta| < 1.57$  
is excluded as it leads to lower-quality 
reconstructed clusters, due mainly to 
services and cables exiting between the barrel and endcap calorimeters.
Electron tracks are reconstructed using an
algorithm~\cite{GSF} that accounts for possible energy loss due to
bremsstrahlung in the tracker layers.  
The energy of an electron candidate with $\et>20~\gev$ is essentially 
determined by the ECAL cluster energy, while its momentum direction 
is determined by that of the associated track.
Particles misidentified as
electrons are suppressed by requiring that the $\eta$ and $\phi$ coordinates
of the track trajectory extrapolated to the ECAL match the $\eta$ and
$\phi$ coordinates of the ECAL cluster, by requiring a narrow ECAL 
cluster width in $\eta$, and by limiting the HCAL energy measured in a 
cone of $\Delta R < 0.15$ around the ECAL cluster direction.
\par
Electrons from photon conversions are suppressed by requiring 
one hit in the innermost pixel layer
for the
reconstructed electron track.  Furthermore, electrons are
rejected when a partner track is found that is consistent with a
photon conversion, based on the opening angle and the separation in 
the transverse plane at the point at which the electron and partner
tracks are parallel.
\par
For both the $\PW$ and $\PZ$ analyses
an electron candidate is considered isolated if $\ITRK <0.09$, $\IECAL < 0.07$
and $\IHCAL < 0.10$ in the barrel region;  $\ITRK <0.04$, $\IECAL < 0.05$
and $\IHCAL < 0.025$ in the endcap regions.
\par
The electron selection criteria were obtained
by optimizing signal and background levels according to 
simulation-based studies. The optimization was done for EB
and EE separately.  We use the same criteria for the $\Wen$ and 
$\Zee$ channels; these select approximately 75\% of the 
reconstructed electrons in the data with clusters in the ECAL fiducial 
volume and 
$\et>20~\GeV$, and reduce the 
fake electron background by 
two orders of magnitude.
\par
More details and studies of electron reconstruction and identification 
can be found in Ref.~\cite{CMS-PAS-EGM-10-004}.

\subsection{Muons}
\par
Events with high-$\pt$ muons are selected online if the data from the muon chambers
satisfy the L1 muon trigger, and if a muon candidate
reconstructed from both muon chamber and tracker data satisfies
the HLT.   An HLT threshold of $\pt > 9 \GeV$ for muons in the 
range $|\eta|<2.1$ is chosen as the baseline for the analysis. 
\par
Offline, a number of quality requirements are imposed.
Muon candidates can be reconstructed by two different 
algorithms: one starts from inner-tracker information 
(``tracker muons''), and another starts from segments 
in the muon chambers  (``global muons''). We demand that muon
candidates for this analysis be reconstructed by both algorithms.
We also demand signals in at least two muon stations, and
require that $\chi^2/{N_{\mathrm{dof}}} < 10$
for a global fit 
containing all valid tracker and muon hits,
where $N_{\mathrm{dof}}$ 
is the number of degrees of freedom.  
The first condition
ensures a sensible momentum estimate at the muon trigger level, 
and further suppresses remaining punch-through and sail-through
hadrons.  
The second condition suppresses contributions from
light-meson decays-in-flight.
\par
In order to ensure a precise estimate of momentum and impact parameter, 
only tracks with more than $10$~hits in the tracker and at least one hit 
in the pixel detector are used. Cosmic-ray muons are rejected by requiring 
an impact parameter relative to the nominal beam 
axis of less than $2$~mm. 
Studies of cosmic-ray muons
confirm that the high-$\pt$ cosmic 
muon contamination is negligible.
\par
As in the case of electrons, isolation criteria are applied.
For both $\PW$ and $\PZ$ analyses, a muon candidate is considered 
isolated if $\IRelComb < 0.15$.
\par
More details and studies of muon reconstruction and identification 
can be found in Ref.~\cite{CMS-PAS-MUO-10-002}.

\section{Missing Transverse Energy}
\label{sec:MET}
\par
An accurate $\MET$ measurement is essential for distinguishing
a $\PW$ signal from QCD multijet production backgrounds.  
We profit from the application
of the particle flow (PF) algorithm~\cite{CMS-PAS-PFT-10-003}, which provides superior $\MET$ reconstruction performance at
the energy scale of $\PW$ boson production. 
The algorithm combines information from
the inner tracker, the muon chambers, and all the calorimetry cells
to classify reconstructed objects according to particle type
(electron, muon, photon, charged or neutral hadron),
thereby allowing precise energy corrections, and
also providing a significant degree of redundancy that 
reduces the sensitivity of the $\MET$ measurements 
to miscalibrations of the calorimetry.
\par
Anomalous noise signals can spoil the $\MET$ measurements.
A dedicated effort to identify and remove such noise in
the ECAL and HCAL, based on cosmic-ray and control
samples as well as collision data, has successfully 
reduced the impact of such noise to a negligible level;
there is no discernible difference in the $\MET$
distributions for $\Wln$ events from data and from
simulation~\cite{CMS-PAS-JME-10-005}.
\par
The $\MET$ is the modulus of the transverse missing momentum 
vector, computed as the 
negative of the vector sum of all 
reconstructed transverse momenta of particles identified with the PF algorithm.
The $\MET$ resolution for inclusive multijet samples and for
$\Wln$ events is reproduced well by the 
simulation.  The resolution worsens by about $10\%$  
when there is more than one primary vertex; this occurs in  
about $40\%$ of the events in the considered data set, 
and has a negligible impact on the
extraction of the $\PW$ signal yields described below.
\section{Lepton Selection Efficiencies}\label{sec:efficiencies}
\par
The efficiencies for lepton reconstruction, identification, 
isolation and trigger
efficiencies 
are obtained from data.  
Correction factors for the values extracted from the simulation
are determined 
with a tag-and-probe method exercised on
$\Zll$ samples in both data and simulation.
This procedure
adequately removes any systematic uncertainties coming from imperfections in 
the simulation, even though the kinematic 
distributions
of leptons in the $\Zll$ sample
differ
slightly from 
those 
in the selected $\Wln$ sample.
\par
The tag-and-probe sample for the measurement of a given efficiency 
contains events selected with two lepton candidates.
One lepton candidate, called the ``tag,'' satisfies
tight identification and isolation requirements. The other 
lepton candidate, called the ``probe,'' is selected with 
criteria that depend on the efficiency being measured.
The invariant mass of the tag and probe lepton candidates 
must fall in the range $60$--$120~\GeV$.
The signal yields are obtained for two exclusive subsamples of events 
in which the probe lepton passes or fails the selection criteria considered.
Fits are performed to the invariant-mass distributions
of the pass and fail subsamples, including a term that
accounts for the background. 
The measured efficiency is  
deduced from the relative level of signal in the pass and fail subsamples;
its uncertainty includes a systematic contribution from the fitting
procedure.  
\par
The correction factors are obtained as
ratios of tag-and-probe efficiencies for
the data and for the simulation.  They are used to compute
the signal selection efficiency ratios $\RHO{}$, and
their uncertainties are  propagated as systematic uncertainties 
on these quantities, 
except in the $\Zmm$ analysis, for which the efficiencies and yields
are determined simultaneously.
\par
The efficiency of the lepton isolation requirements
can also be measured using a ``random-cone'' technique.
In the inclusive $\PW$ or $\PZ$ sample, energy contributing to
the isolation variables comes mainly from the underlying
event, which can be sampled in directions uncorrelated
with the lepton directions in a particular event.  We
use leptons in simulated signal events to define
directions in data events where the isolation energies
can be measured and compared to the requirements of the
analysis; this ensures a sampling of phase space that
mimics the leptons in real data events.  Studies with
simulation verify that this technique provides values
for the isolation efficiency that are accurate to
about 0.5\% for muons and 1\% for electrons.

\subsection{Electrons}
\par
The electron selection efficiency is the product of three components:~1)~the 
reconstruction efficiency,~2)~the 
identification and isolation
efficiency,
and~3)~the trigger efficiency.  Efficiencies are evaluated 
for the barrel and endcap regions, and for the two possible
electron charges,  separately.
\par
The reconstruction efficiency is the probability of finding a reconstructed
track when the electron cluster is within the ECAL fiducial volume.
The probe is selected as an ECAL cluster
of reconstructed transverse energy 
greater 
than $20~\GeV$.  
To reduce backgrounds, which are not 
insignificant,
we use a tight selection on the tag 
and require the probe 
to pass additional loose shower shape and isolation requirements; 
these are known from simulations to be uncorrelated 
with the reconstruction efficiency.
The measured reconstruction efficiency is the fraction of 
probes reconstructed as electron tracks.  
For the EB and EE electrons we measure a reconstruction efficiency 
of $\WPWIEBEFFRECO$
and
$\WPWIEEEFFRECO$, respectively. 
The resulting correction factors are consistent with unity.
\par
The efficiency of electron identification, isolation, and conversion
rejection requirements is estimated relative to the sample of 
reconstructed electrons. The tag selection does not need to be tight,
and no additional criteria on the probe are imposed. 
In the barrel, we measure a selection efficiency of 
$\WPWIEBEFFID$, to be compared to  $\WPWIEBMCID$ for
the simulation, resulting in a correction factor of
$\WPWIEBRID$. In the endcaps, an efficiency of $\WPWIEEEFFID$ 
is 
measured,
where $\WPWIEEMCID$ is expected from simulation, resulting in a
correction factor of $\WPWIEERID$. 
The random-cone technique is used to cross check the efficiency of
the electron isolation requirements.  The results confirm the values
within $1.0\%$ for EB 
and $1.8\%$ for EE electrons, respectively. 
\par
Finally, we obtain combined L1 and HLT trigger efficiencies 
from identified and isolated electron candidates as probes. 
We measure $\WPWIEBEFFHLT$ in the barrel, 
and $\WPWIEEEFFHLT$ in the endcaps, 
leading to correction factors consistent with unity. 
These tag-and-probe efficiencies are confirmed by measurements
made with a sample of minimum-bias events selected with scintillation
counters and a sample of events selected by an HLT algorithm that has 
minimum-bias requirements at L1 and a complete emulation of the 
offline ECAL cluster reconstruction.
\par
The charge misidentification 
for electrons in the simulated $\PW$ sample
is $\EFFA{0.67}{0.01}$. 
We infer a data/simulation
charge misidentification correction factor of $\WPWMISID$
by comparing the fraction of events with electrons of same electric charge
in data and simulation samples.
This correction factor is included in the results for $\PW^{\pm}$ 
cross sections, as well as their ratio, and its error propagated to the
systematic uncertainties on these quantities. 
\par
The products of all correction factors for the electron selection
are $\WPWIEBR$ for the EB and $\WPWIEER$ for the EE.
\par
When combining the correction factors, we take into account the relative
acceptance of electrons from $\PW$ decays in the EB and EE.
We obtain the efficiency ratio  for  $\Wen$ events: $\RHO{\PW} = \WEITNPR$;
and separately by charge: $\RHO{\PWp}=\WEPTNPR$ and $\RHO{\PWm}=\WEMTNPR$.
We infer a signal selection efficiency of $\WEIEFF$ for 
$\Wen$ events with the electron cluster in the ECAL fiducial volume 
and $\et>20~\GeV$.

In the $\Zee$ analysis, one electron 
candidate is allowed to fail the trigger criteria;
the efficiency ratio is $\RHO{\PZ} = \ZEETNPR$ and
the corrected signal selection efficiency for $\Zee$ events with 
both electron clusters in ECAL fiducial volume and $\et>20~\GeV$ 
is $\ZEEEFF$.  
This number is derived from the corrected overall electron 
selection efficiencies, which are $\WPWIEBSELEFF$ and $\WPWIEESELEFF$ in the EB and EE, respectively, and taking into account the expected fractions of $\Zee$ events with 
EB-EB, EB-EE and EE-EE combinations of electrons, which are
$\ZEBEBFRAC$, $\ZEBEEFRAC$ and $\ZEEEEFRAC$, respectively.

\subsection{Muons}
\par
The muon reconstruction and selection efficiency has five 
distinguishable components:~1)~the efficiency to find a track 
in the inner tracker,~2)~the efficiency to find a track in the muon chambers,
and, for a muon candidate,
3)~the efficiency to pass the quality requirements, 
4)~the efficiency to pass the isolation requirements, and 
5)~the probability to pass the L1 trigger and HLT.
\par
Muon efficiencies are extracted from the sample 
of candidate $\Zmm$ events. The tag muon
must pass all muon selection criteria. 
The invariant mass 
of the tag-and-probe muon candidates is formed; invariant-mass
distributions are produced for
exclusive categories of events where the probe muon
passes or fails various efficiency requirements.
Simultaneous fits to those distributions allow the number of 
signal events and the efficiencies to be extracted.  
\par
The inner-tracker efficiency is studied using well-reconstructed 
tracks in the muon chambers as probes. The efficiency for tracking 
in the muon chambers is tested with tracker muons satisfying very 
loose matching to muon track segments. To measure the efficiency 
of quality requirements, the probe muon must pass all the selection criteria
except those on the $\chi^{2}$ and on the impact distance to the 
beam 
axis.
Finally, the isolation efficiency is measured using 
muons that pass the quality requirements, and the trigger efficiency
using muons that in addition are isolated.
\par
The following efficiencies are obtained: for inner tracking, $\WMUIEFFTRK$; 
for muon tracking, $\WMUIEFFSA$; for quality requirements, $\WMUIEFFSEL$; for
isolation, $\WMUIEFFISO$; and for trigger, $\WMUIEFFTRG$.  All
correction factors are 
consistent with unity, except for the trigger efficiency, for which the 
correction factor is $\WMUIRTRG$.  
\par
Isolation efficiencies have also been measured using the random-cone 
technique, and the results confirm the tag-and-probe value for the 
isolation efficiency quoted above: $98.7\%$ for $\Wmn$ and 
$98.5\%$ for $\Zmm$.
\par
The overall muon selection efficiency is
$\WMUIEFF$, to be compared to the value of $\WMUIMCEFF$ obtained from the 
simulation; the efficiency ratio is $\RHO{\PW} = \WMUIR$.
There is no significant difference
between the efficiency ratios for positive and negative
muons: 
$\RHO{\PWp}=\WMIEFFPLS$ and $\RHO{\PWm}=\WMIEFFMIN$, respectively.
\section{Event Selection and Signal Extraction}
\label{sec:selection}
\par
The data used for these measurements 
were collected from April to August 2010.  
We used only those data-taking periods passing the standard CMS
quality criteria, which allow no anomalous or
faulty behavior for the inner tracker, the calorimeters,
and the muon chambers.  
\par
Several large samples of simulated events were used to evaluate the signal
and background efficiencies and to validate our analysis techniques.
Samples of electroweak processes with $\PW$ and $\PZ$~production, both for
signal and background events, were produced with 
POWHEG~\cite{Alioli:2008gx, Nason:2004rx, Frixione:2007vw},
interfaced with the PYTHIA~\cite{Sjostrand:2006za} parton-shower generator. 
QCD events with muons, electrons, or jets likely to be misidentified 
as electrons in the
final state were studied with PYTHIA, as were other minor
backgrounds such as $\ttbar$ and certain electroweak processes
($\Wtn$, $\Ztt$, $\WW$, $\WZ$, and $\ZZ$). 
We do not consider the diboson channels ($\WW$, $\WZ$, and $\ZZ$) 
as part of the $\PW$ and $\PZ$ signals
in order to facilitate the comparison of our results to theoretical predictions,
which do not take these contributions into account. 
Generated events were
processed through the full GEANT4~\cite{GEANT4} detector simulation,
trigger emulation, and event reconstruction chain.

\subsection{$\PW$ boson selection}

\par
The $\PW$ events are characterized by a prompt, energetic, and
isolated lepton, and significant missing energy.  
The main backgrounds are QCD multijet events and Drell-Yan
events in which one lepton fails the selection.  The QCD background is
reduced by requiring the lepton to be isolated; the remaining
events do not have large $\MET$ and can be distinguished from
signal events on a statistical basis.  The Drell-Yan background
is suppressed by rejecting events with a second lepton candidate.

To measure the signal yields, we choose to fit the $\MET$ 
distribution in the electron channel and the $\MT$ distribution
in the muon channel, where
$\MT=\sqrt{2\pt \MET (1-\cos{\Delta\phi})}$;
$\Delta\phi$ is the angle between the 
missing transverse momentum and the lepton transverse momentum.  
QCD backgrounds are estimated from data, as explained below.
According to the simulation,
$\Wtn$ makes a
small relative contribution;
backgrounds from $\Ztt$, $\ttbar$, and diboson production 
are negligible in both electron and muon channels.

\subsubsection{Electrons}

The $\Wen$ candidate events are required to have one identified electron 
with an ECAL cluster of $\et > 20\GeV$ in the ECAL fiducial volume. 
If a second electron candidate 
satisfying looser criteria and with $\et > 20\GeV$ is present in
the event, the event is rejected.  
The fraction of signal events selected in the simulation is
$\APRIM{\PW} = \WEIAPRIM$, with  $\APRIM{\PWp}=\WEPAPRIM$ and 
$\APRIM{\PWm}=\WEMAPRIM$.
The number of events selected in the data
is $\WEISAMPLE$, with $\WEPSAMPLE$ positive and $\WEMSAMPLE$ 
negative electrons.
\par
The $\Wen$ signal is extracted from an unbinned maximum likelihood 
fit of the observed $\MET$ distribution to the sum of signal and
background shapes.
The QCD background shape, which accounts
for both QCD multijet production and direct-photon
production with the photon converting in the detector,
can be modeled by a modified Rayleigh distribution,
$$f(\MET) = \MET \times \exp{\left(-\frac{\MET^2}{2(\sigma_0+\sigma_1\MET)^2}\right)}.$$ 
This function can be understood as describing fluctuations
of the 
missing transverse momentum
vector around zero due to measurement errors;
the resolution term, 
$\sigma_0+\sigma_1 \MET$, 
increases
with $\MET$ to account for tails in the $\MET$
measurement.  This function describes well the QCD
background shape in the simulation, over the full
range of $\MET$, as well as $\MET$ distributions from
signal-free samples obtained by inverting the
identification or isolation criteria. 

The signal distributions are derived from simulation, 
separately for $\PWp$ and $\PWm$,
and receive
an event-by-event correction in bins of the $\PW$ transverse momentum, 
determined from a study of the hadronic recoil 
distributions of $\Zee$ events in the data~\cite{CMS-PAS-JME-10-005}.
In fits to the $\MET$ distributions, the free parameters 
are the $\PW$ signal yield, the QCD background yield, and 
the shape parameters $\sigma_0$ and $\sigma_1$.  

We extract the inclusive yield $\NSIG{\PW}$ from a fit where
the expected ratio for $\XVL{\sigma}{\PWp}/\XVL{\sigma}{\PWm}$ 
is assumed. 
It has been  checked that the result was insensitive to this assumption. 
Figure~\ref{fig:W}~(a) shows the $\MET$ distribution of 
the inclusive $\Wen$ sample
and the results of the likelihood fit;  the fit function describes the 
data well, with a $p$-value of $\WEIKSPCOR$ for the Kolmogorov-Smirnov test.
The inclusive yield is $\NSIG{\PW} = \WEIYIELD$ events.

The signals for the $\Wpen$ and $\Wmen$ channels are 
extracted from a simultaneous fit to the individual $\MET$
distributions, in which the QCD background  shape parameters $\sigma_0$ and
$\sigma_1$ are constrained to be the same for both samples. 
The yields are $\NSIG{\PWp}=\WEPYIELD$ for $\Wpen$ and $\NSIG{\PWm}=\WEMYIELD$ 
for $\Wmen$, with a negligible correlation.
Because the two fits are independent, 
the relation $\NSIG{\PW} = \NSIG{\PWp} + \NSIG{\PWm}$ is not exactly satisfied, 
but holds to within 0.2\%. 

\subsubsection{Muons}
\par
The $\Wmn$ candidate events are required to have a muon 
with $\pt > 20~\GeV$ and $|\eta|<2.1$.   If a second muon
with $\pt > 10~\GeV$ is present, the event is rejected
in order to reduce the contribution from Drell-Yan events.
The fraction of signal events selected from the simulation is
$\APRIM{\PW} = \WMIAPRIM$, with  $\APRIM{\PWp}=\WMPAPRIM$ and 
$\APRIM{\PWm}=\WMMAPRIM$.
The number of selected events is $\WMISAMPLE$, including $\WMPSAMPLE$ 
with positive and $\WMMSAMPLE$ with negative muons.
\par  
The $\Wmn$ signal yield is extracted from a binned
likelihood fit to the observed $\MT$ distribution,
which is taken to be the sum of different contributions:
$\Wmn$ signal, QCD background, electroweak (EWK) backgrounds,
and $\ttbar$. The shapes of the signal and background components 
(templates) are taken from the simulation, except for the
QCD background, which is obtained from data, as 
described below.
The normalization of the QCD background and the $\Wmn$ yield are free
parameters in the fit.  The EWK and  $\ttbar$  backgrounds are 
normalized to  the $\Wmn$ yield on the basis of simulations 
and expected relative cross sections.
\par
The QCD template used in the fit is obtained from a
high-purity QCD sample referred to as the inverted sample. 
This sample is selected by applying the same criteria as in the 
signal selection except the isolation requirement, which is reversed:
$\IRelComb > 0.20$. The shape of the QCD template 
from the inverted sample in the data 
agrees well with that obtained in the simulation. 
Studies of simulated QCD events show that
a small bias in the shape is induced by the isolation requirement.  
This bias comes from the correlation of the isolation variable
with the $\sum \et$ in the event.  We correct the template on
the basis of a linear relation between $\MT$ and $\IRelComb$. In the 
simulation, we obtain an excellent match between the corrected 
template from the inverted sample and the actual template from the 
non-inverted sample.
We compare the yields obtained when fitting with different QCD templates,
namely, corrected template in the data and uncorrected templates
obtained from the inverted sample in the data and from the non-inverted
sample in the simulation. We take the 
maximum difference in yields as an estimate of the systematic uncertainty 
from the modeling of the QCD background shape. 
\par
As in the case of electrons, the signal template
receives an event-by-event correction in bins of the $\PW$ transverse momentum 
determined from a study of the hadronic recoil 
distributions of $\Zmm$ events in the data.
\par
\begin{figure}[t]
\begin{center}
         \includegraphics[width=.45\textwidth]{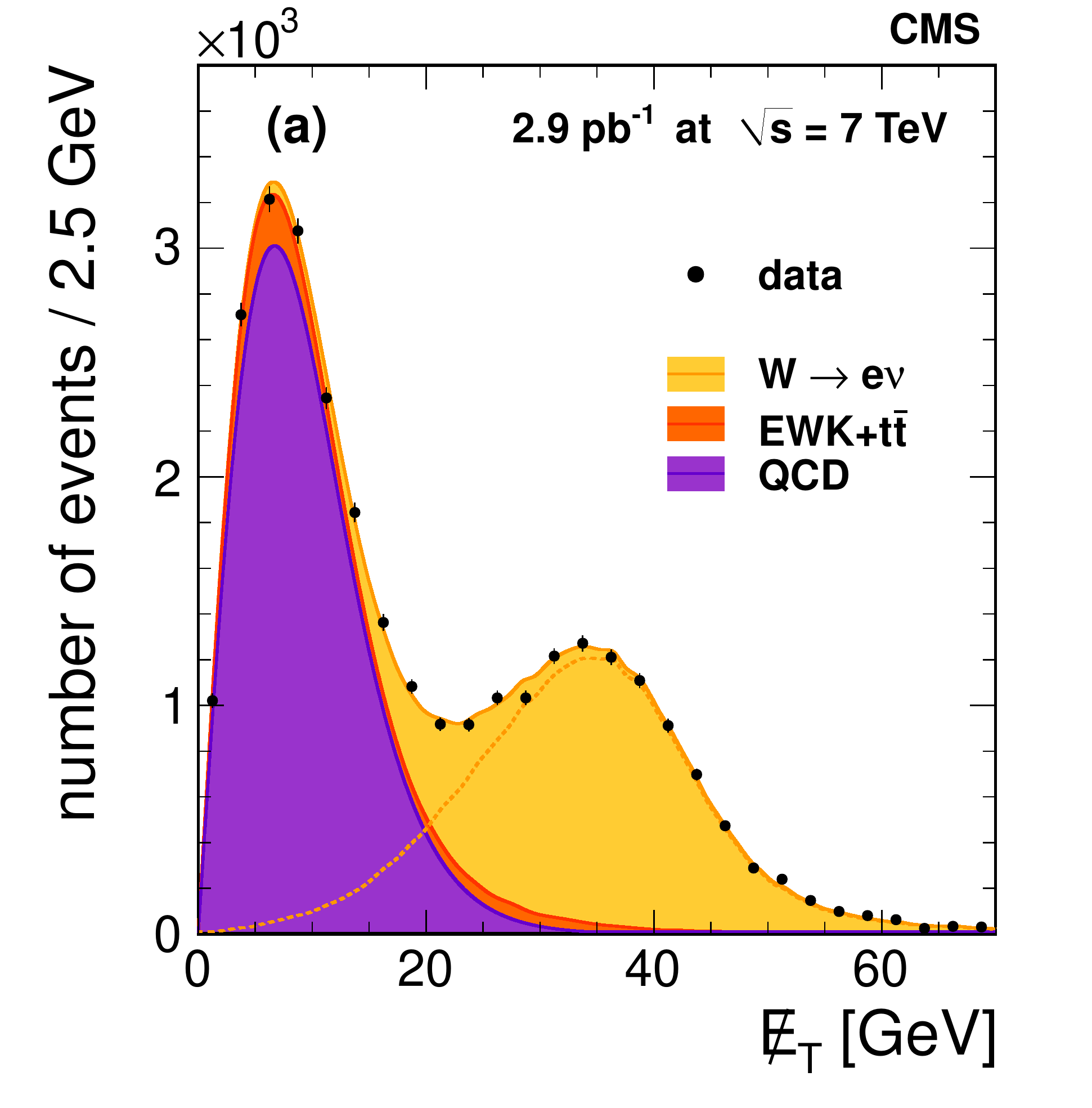} 
        \hspace{.05in}
         \includegraphics[width=.45\textwidth]{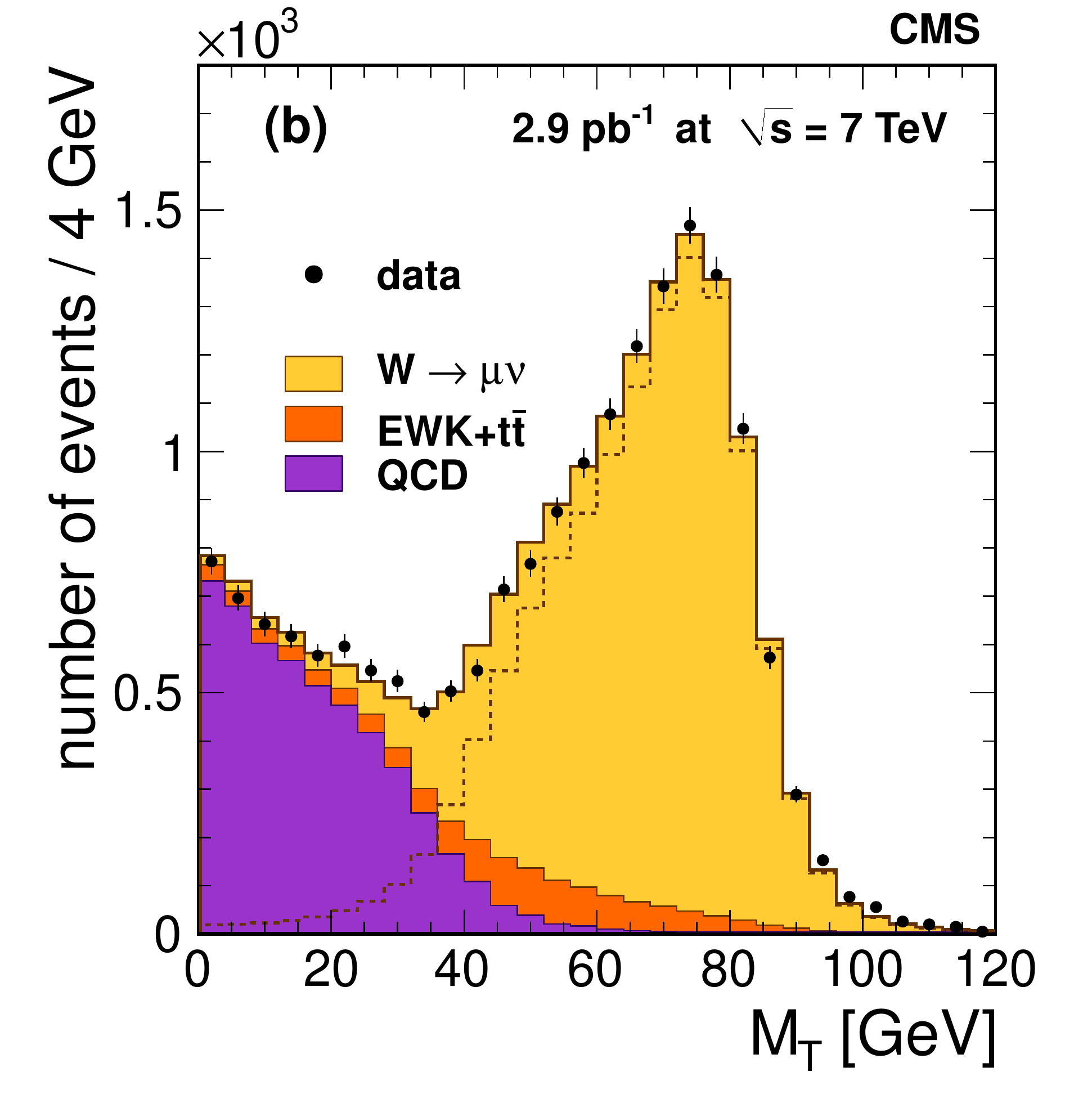} 
        \hspace{.05in}
       \caption{The $\PW$ signal distributions:~(a)~$\MET$ 
distribution for the selected $\Wen$ sample;~(b)~$\MT$ 
distributions for the selected $\Wmn$ sample.  The points represent
the data.  Superimposed are the results of the maximum likelihood fits for
signal plus backgrounds, in yellow; 
all backgrounds, in orange;
QCD backgrounds, in violet.
The dashed lines represent the signal distributions.
\label{fig:W} }
\end{center}
\end{figure}

Figure~\ref{fig:W}~(b) shows the fit to the observed $\MT$ spectrum
of the inclusive $\Wmn$ sample; the fit distribution describes the 
data well, with a $p$-value of $\WMIKSPCOR$ for the Kolmogorov-Smirnov test.
The inclusive yield is $N_{\PW} = \WMIYIELD$. 
The charge-specific yields are
$N_{\PWp}=\WMPYIELD$ 
and $N_{\PWm}=\WMMYIELD$. 
Here, we fit simultaneously for the inclusive yield $N_{\PW}$ 
and the ratio $N_{\PWp}/N_{\PWm}$ so that,
by construction, $N_{\PW} = N_{\PWp} + N_{\PWm}$. 

\subsection{$\PZ$ boson selection}
\par
To identify $\Zll$ decays,
a pair of identified leptons is required, with dilepton invariant
mass in the range $60 < \MLL < 120\GeV$.
Backgrounds are very low, including backgrounds from QCD processes.
In the $\Zee$ channel, the yield is obtained by counting the number 
of selected events and making a small correction for backgrounds.
In the $\Zmm$ channel, 
yield and lepton efficiencies are fitted simultaneously.
No correction is made for $\gamma^*$ exchange.

\subsubsection{Electrons} \label{sec:Z-electrons}
\par
The $\Zee$ candidate events are required to have two electrons satisfying
the same selection criteria as the electrons selected
in the $\Wen$ sample.  Both electrons must have an ECAL cluster with 
$\et > 20$~GeV in the ECAL fiducial volume.  
The fraction of signal events selected in the simulation is
$\APRIM{Z}=\ZEEAPRIM$.
\par
The $\PZ$ mass 
peaks
in the data 
exhibit small shifts, on the order of 1 to 2\%,
with respect to the simulated distributions.
From these shifts, we determine 
ECAL cluster energy 
scale correction factors 
of \EBESCALE and \EEESCALE
for barrel and endcap electrons, respectively. 
The uncertainties on these correction factors are
propagated as systematic uncertainties on the yield.
Applying these corrections to 
electron candidates in the data, we 
select $\ZEESAMPLEN$ events, with the dielectron invariant mass
shown in Fig.~\ref{fig:Z}~(a),
along with the predicted distribution,
after the energy scale correction of the data and normalization of the 
simulation.
\par
Three techniques are used to estimate
the background originating from events in which
one or both electron candidates are misidentified jets or photons. 
The first method measures the probability of jets to be misidentified as 
electron from a large sample of events selected with a jet trigger. 
The second method is based on counting events with 
electron candidates of same electric charge, after taking into account the 
probability of wrong charge assignment. The third method uses 
a fit to the track isolation variable to extract the 
fractions of signal and QCD background. The three methods are independent
and give consistent results. Combining them, 
we estimate the QCD background in our sample to be $\ZEEQCDBKG$ events.
Backgrounds from other processes with true electrons ($\Ztt$, dibosons, and
$\ttbar$) are estimated from the simulation. The total 
background in the $\Zee$ sample is estimated to be \ZEEBKG events.

\subsubsection{Muons}
\par
\begin{figure}[t]
\begin{center}
         \includegraphics[width=.45\textwidth]{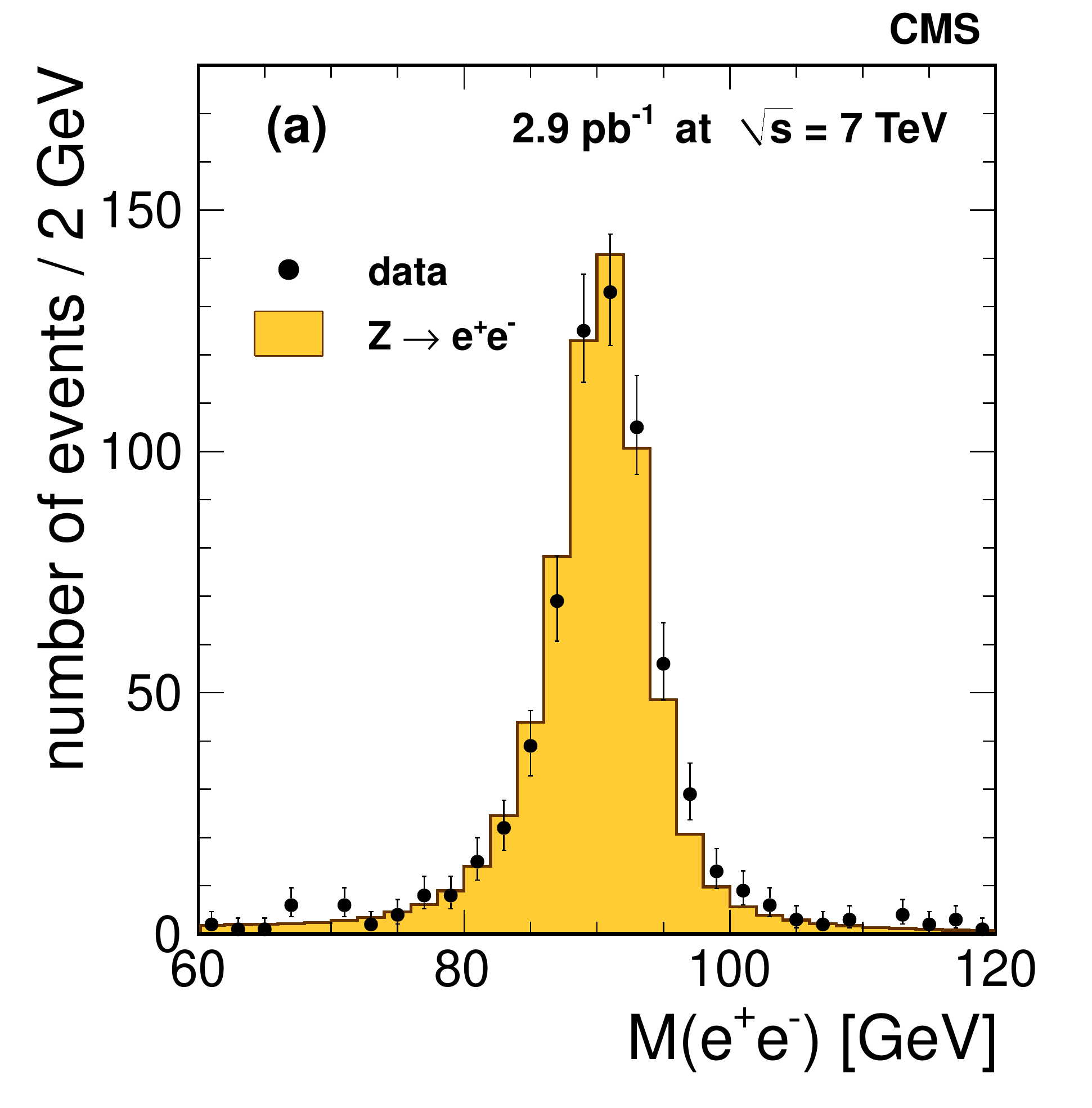} 
        \hspace{.05in}
         \includegraphics[width=.45\textwidth]{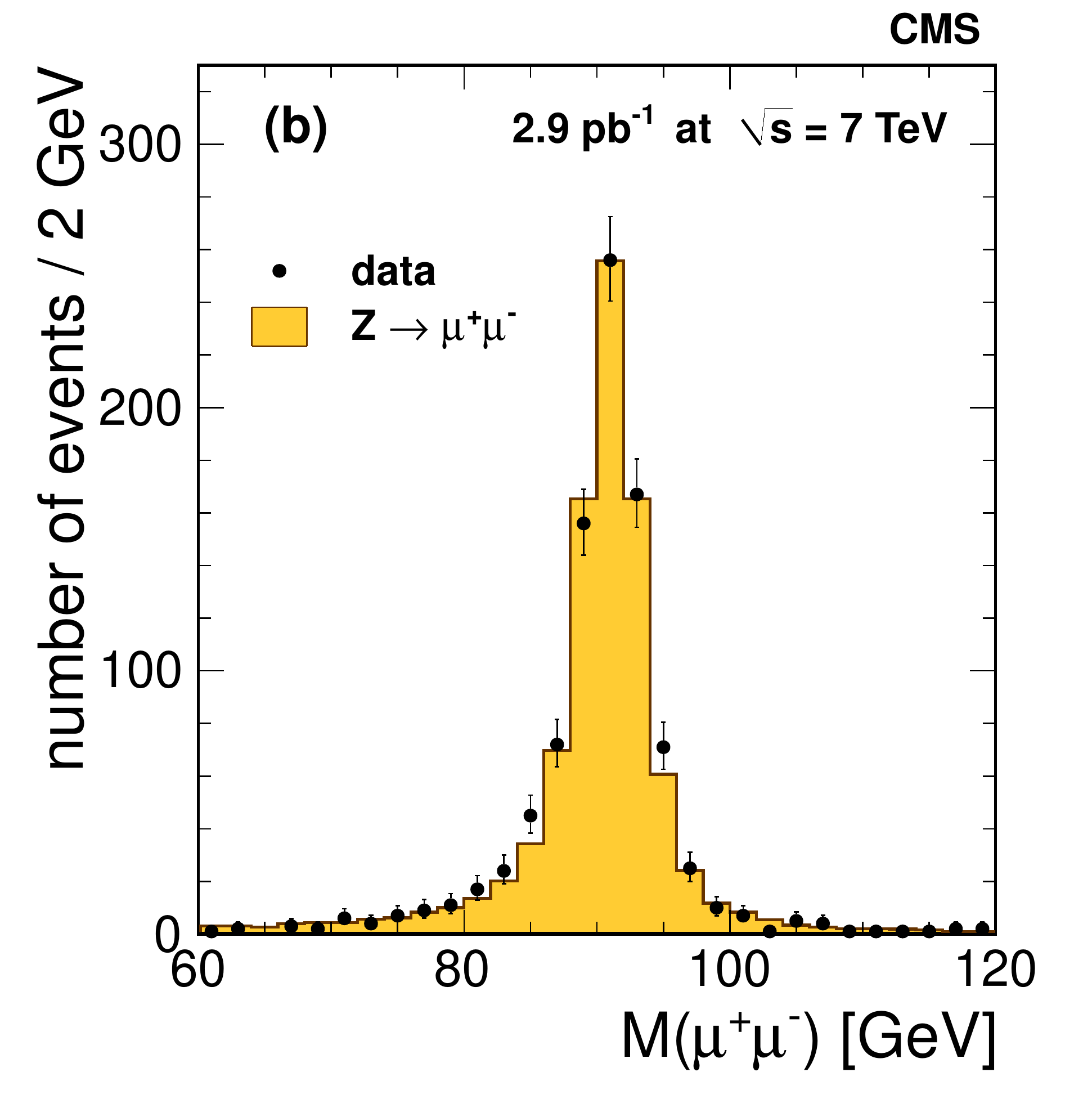} 
        \hspace{.05in}
       \caption{The $\PZ$ signal distributions:~(a)~dielectron 
mass spectrum for the selected $\Zee$ sample;~(b)~dimuon mass 
spectrum for the selected $\Zmm$ sample.
The points represent the data and the histograms, the simulation. 
Backgrounds are negligible and are not
represented in the plots.
\label{fig:Z} }
\end{center}
\end{figure}

In the $\Zmm$ channel, event yields and muon selection efficiencies 
are extracted from a simultaneous fit.
The tag-and-probe sample is built from events containing
two muon candidates with $\pt > 20~\GeV$  
and $|\eta| < 2.1$. The tag muon
satisfies the identification and isolation criteria used 
in the $\Wmn$ selection; the probe muon is selected
as either a tracker or global muon. The tag-and-probe sample 
is divided into five mutually-exclusive samples of events, 
according to the quality of the probe muon, as described above.
In the signal sample, the probe muon fulfills all 
the identification and isolation criteria, and
at least one of the muon candidates satisfies the 
trigger requirement. This sample contains $\ZMMSAMPLE$ events.  
The distribution of the
dimuon invariant mass is shown in Fig.~\ref{fig:Z}~(b), 
compared with distributions based on simulations 
normalized to the measured cross section.  
\par
The background is negligible in the signal sample.
The mass spectrum in that sample is used as a  model for the signal 
shapes in other
samples, where backgrounds are modeled by products of a polynomial 
and an exponential
function. The yields and efficiencies are extracted from a joint 
binned maximum likelihood fit to all mass spectra.
The $\Zmm$ signal yield is already corrected 
for efficiency by virtue of the parameterization used in the fit; 
the corrected yield is $\NSIG{\PZ}/\EPSB{\PZ}=\ZMMYIELD$ events 
and the signal acceptance is $\AGEN{\PZ}=\ZMMAGEN$.
\par 
The muon momentum scale and resolution are 
verified
in different $\pt$ regions
from the study of lower-mass dimuon resonances ($\PJgy$ and
$\PgU$), the cosmic-ray muon endpoint~\cite{MuonChargeRatio}, the matching of tracker muons and
global muons, the $\PW$ transverse momentum spectrum, 
and the $\PZ$ mass lineshape.  
From the observed agreements with the simulation, we find that no momentum 
corrections are needed.
\par
The QCD multijet background in the signal sample is estimated to be
$0.048\pm0.002$ event. 
Including or neglecting this background in the
simultaneous fit changes the yield by 0.2\%, which
we take as a systematic uncertainty on the background.
A further systematic uncertainty stems from the modeling
of the shapes of signal and background; we estimate
this uncertainty to be 1\%.  The contributions from
other backgrounds ($\Ztt$, dibosons, and $\ttbar$)
are small, according to simulations, and amount to
$3.5\pm 0.2$ events in total. 
\section{Systematic Uncertainties \label{sec:systematics}}

\par
The largest uncertainty for the cross section measurement comes from
the estimation of the integrated luminosity. 
CMS uses signals from the forward hadronic (HF) 
calorimeters to measure
the instantaneous luminosity in real time with an absolute
normalization obtained with Van der Meer scans, 
from which we infer the size of
the colliding beams and thereby the luminosity, 
with minimal reliance on simulations~\cite{CMS-PAS-EWK-10-004}.
The total luminosity uncertainty amounts to
$11\%$ and is expected to diminish in the future.
 
\par
Aside from luminosity, the main source of experimental uncertainty 
in our measurements comes from the propagation of 
uncertainties on the efficiency ratios
obtained by the tag-and-probe method. This amounts to $\WEITNPSYST\%$ 
and $\WMIEFFSYST\%$ for $\Wen$ and $\Wmn$ analyses, respectively.
In the $\Zee$ channel, we conservatively neglect
the anti-correlation between efficiencies and yields, 
which are extracted separately from the same sample; the efficiency 
uncertainties amount to $\ZEETNPSYST\%$.
In the $\Zmm$ analysis,
yield and efficiencies are determined simultaneously, and therefore the
efficiency uncertainties are part of the statistical error from the fit.
Corrections of 0.5\% and 1.0\% are applied to the $\Wmn$ and $\Zmm$
event yields, respectively, to account for a loss of events due 
to barrel muon triggers 
that failed timing requirements and for which the tracker data were
not read out properly.
These corrections are determined from the data, 
and lead to a $0.5\%$ systematic uncertainty in both channels.
\par
Sub-dominant systematic uncertainties come from the lepton 
energy/momentum scale and resolution.
Electron energy correction factors are
approximately $1\%$ to $3\%$ in the barrel and endcap calorimeters,
from the observed shift of the $\PZ$ mass peak.
In the $\Wen$ case, the electron energy scale has an impact on the 
\MET distribution for the signal;
we apply typical energy scale corrections to electrons
in the simulation (before $\et$ threshold selection) and
recompute the $\MET$. From variations of the signal yields 
from the fit, we assign a $\WEIESCALESYST\%$ systematic uncertainty 
to the $\Wen$ cross section.  In the $\Zee$ analysis,
the $\et$ threshold and mass window requirements lead to a 
$\ZEEESCALESYST\%$ uncertainty due to the energy scale uncertainty.  
Studies of the $\Zmm$ line shape  
show that data/simulation momentum scale shifts larger than $0.4\%$ 
can be excluded, 
which imply small
uncertainties of $\WMISCALESYST\%$
in the $\Wmn$ analysis, and $\ZMMSCALESYST\%$ in the $\Zmm$ analysis.
\par
The \MET energy scale is affected by our limited knowledge 
of the intrinsic hadronic recoil response.
We observe minor discrepancies when comparing
hadronic recoil distributions in data and simulation,
and assign an uncertainty of  $\WEIMETSYST\%$ in the $\Wen$
analysis due to the $\MET$ energy scale.
In the muon channel, this uncertainty is estimated by refitting the $\MT$ 
distribution with the signal shape predicted by the simulation. 
The variation in the signal yield with respect 
to the reference result is $\WMIMETSYST\%$.
\par
In the $\Wen$ channel, the systematic uncertainty 
due to background subtraction is obtained by comparing 
fits to various background-dominated distributions:
the sample selected with inverted identification criteria 
in the data, and the 
samples selected
with and without inverted identification criteria in the QCD simulation. 
We quantify 
the differences in the tails of these three distributions 
by an extra parameter in our analytical background function.
Using a set of pseudo-experiments to estimate
the impact of such differences on the results of the nominal fit, we
evaluate the uncertainty due to background subtraction in the $\Wen$
analysis to be $\WEIBKGSYST\%$. In the $\Wmn$ analysis
the QCD background shape is tested by refitting the $\MT$ spectrum 
with the background shape fixed to QCD-enriched sample expectations. This 
choice provides the maximum variation ($\WMIQCDSHAPESYST\%$) 
in the signal yield with respect to the reference fit.
\par 
The background from 
fake electrons in the $\Zee$ sample 
is estimated from data, as described in Sect.~\ref{sec:Z-electrons}.  
The uncertainty on this
background is $\ZEEBKGSYST\%$ of the total $\PZ$ yield. 
The expected background to $\Zmm$ is~0.5\%, with an
uncertainty of~0.2\%.   Further uncertainty arises from 
the fit model of the backgrounds in subsamples where 
one of the muon candidates fails the selection. We 
estimate this uncertainty to be~1\%.   
Uncertainties from the normalization of electroweak and $\ttbar$ 
backgrounds are negligible in both $\PW$ and $\PZ$ channels. 
\par
Theoretical uncertainties in the $\Wln$ cross section measurement affect
the estimation of the acceptance. 
The Monte Carlo estimates are based on simulations
that use a next-to-leading order (NLO) generator (POWHEG) as input. Events are re-weighted
at generator level according to different PDF set assumptions 
(CTEQ6.6~\cite{Nadolsky:2008zw}, MSTW08NLO~\cite{Martin:2009iq},
NNPDF2.0~\cite{Ball:2010de}). The observed variations in the acceptance 
are less than $1.2\%$. Remaining theoretical uncertainties associated with 
the treatment of
initial-state radiation, final-state QED radiation, missing
electroweak effects, and renormalization and factorization scale assumptions
amount to approximately $1.5\%$.
\par
Table~\ref{tab:syst} shows a summary of the systematic uncertainties 
for the $\PW$ and $\PZ$ cross section measurements.

\begin{table}
   \caption[.]{ \label{tab:syst}
Systematic uncertainties of the four cross section measurements, in percent.
``n/a'' means the source does not apply.
A common luminosity uncertainty of 11\% applies to all channels. 
}
\begin{center}
\begin {tabular} {|l|c|c|c|c|}
\hline
Source       & $\Wen$ & $\Wmn$ & $\Zee$ & $\Zmm$ \\
         \hline
Lepton reconstruction \& identification  & \WEITNPSYST    & \WMIEFFSYSTPRET  & \ZEETNPSYST    & \ZMMEFFPRET \\
Momentum scale \& resolution             & \WEIESCALESYST & \WMISCALESYST    & \ZEEESCALESYST & \ZMMSCALESYST \\
$\MET$ scale \& resolution               & \WEIMETSYST    & \WMIMETSYST      &  n/a           &  n/a   \\
Background subtraction/modeling          & \WEIBKGSYST    & \WMIQCDSHAPESYST & \ZEEBKGSYST    
& $\ZMMBKGSYST \oplus \ZMMFITSYST$   \\
PDF uncertainty for acceptance           & \WEIPDFACCSYST & \WMIPDFACCSYST   & \ZEEPDFACCSYST & \ZMMPDFACCSYST \\
Other theoretical uncertainties          & \WEITHSYST     & \WMITHSYST       & \ZEETHSYST     & \ZMMTHSYST  \\
\hline
Total                                    & \WEITOTSYST    & \WMITOTSYST      & \ZEETOTSYST    & \ZMMTOTSYST \\
\hline
\end {tabular}
\end{center}
\end{table}
\section{Results \label{sec:results}}
\par
All theoretical predictions quoted in this section are computed
at the next-to-next-to-leading order (NNLO) with the program FEWZ~\cite{Melnikov:2006kv, Melnikov:2006di}
and the MSTW08 set of PDFs.  The uncertainties correspond to 68\% confidence
levels obtained by combining the PDF and $\alpha_S$ errors
from the MSTW08, CTEQ6.6, and NNPDF2.0 groups and adding the
NNLO scale uncertainties in quadrature, as prescribed by the
PDF4LHC working group~\cite{PDF4LHC}.

For all measurements we present results for electron and muon 
channels separately and, assuming lepton universality in $\PW$ and $\PZ$ decays, 
for the combined lepton channel.  The electron and muon channels are combined 
by maximizing a likelihood that accounts for the individual 
statistical and systematic uncertainties and their correlations.
For cross section measurements, 
correlations are only numerically relevant for theoretical
uncertainties, including the PDF uncertainties on the acceptance values.
For cross section ratio measurements, the correlations of lepton 
efficiencies are taken into account in each lepton channel, 
with other experimental uncertainties assumed uncorrelated; 
in the combination of lepton channels,  we assume 
fully-correlated uncertainty for the acceptance
factor, with other uncertainties assumed uncorrelated.

\begin{table}[hbt]
\caption[.]{ Summary of production cross section times branching ratio 
measurements and their theoretical predictions.\label{tab:res-xsections} }
\begin{center}
\begin{tabular}{|l|c|r|c|}
\hline
\multicolumn{2}{|c|}{Channel} &  
\multicolumn{1}{c|}{$\sigma \times {\cal B}$  (nb)} & 
\multicolumn{1}{c|}{NNLO (nb)}   \\
\hline
\hline
\multirow{3}{*}{$\PW$} & $\EN$ & \WEISIGBRGHM & \multirow{3}{*}{\THGHMSIGBRWI} \\
                  & $\MN$ & \WMISIGBRGHM & \\
                  & $\LN$ & \WLISIGBRGHM & \\
\hline \hline
\multirow{3}{*}{$\PWp$} & $\EpN$ & \WEPSIGBRGHM & \multirow{3}{*}{\THGHMSIGBRWP} \\
                  & $\MpN$ & \WMPSIGBRGHM & \\
                  & $\LpN$ & \WLPSIGBRGHM & \\
\hline \hline
\multirow{3}{*}{$\PWm$} & $\EmN$ & \WEMSIGBRGHM & \multirow{3}{*}{\THGHMSIGBRWM} \\
                  & $\MmN$ & \WMMSIGBRGHM & \\
                  & $\LmN$ & \WLMSIGBRGHM & \\
\hline \hline
\multirow{3}{*}{$\PZ$} & $\EpEm$ & \ZEESIGBRGHM & \multirow{3}{*}{\THGHMSIGBRZ} \\
                  & $\MM$ &\ZMMSIGBRGHM & \\
                  & $\LpLm$ &\ZLLSIGBRGHM & \\
\hline
\end{tabular}
\end{center}
\end{table}

\par
The measured cross sections times branching ratio for 
$\PW$, $\PWp$, $\PWm$ and $\PZ$ production are reported in 
Table~\ref{tab:res-xsections}, 
for the electron,  muon, and combined lepton ($\ell=\Pe$ or $\Pgm$) 
channels, along with predictions at the NNLO in QCD. 
The reported $\PZ$ boson production
cross sections pertain to the invariant mass range 
$60 < \MLL < 120~\GeV$, and 
are corrected for the fiducial and kinematic acceptance 
but not for $\gamma^*$ exchange.

The ratio of cross sections for $\PW$ and $\PZ$ production is
\begin{eqnarray*}
\RWZ = \frac{\SIGBRSHORT{\PW}}{\SIGBRSHORT{\PZ}} 
= \frac{\NSIG{\PW}}{\NSIG{\PZ}}\frac{\RHO{\PZ}}{\RHO{\PW}}\frac{\APRIM{\PZ}}{\APRIM{\PW}}
= \frac{\NSIG{\PW}}{\NSIG{\PZ}}\frac{\EPSB{\PZ}}{\EPSB{\PW}}\frac{\AGEN{\PZ}}{\AGEN{\PW}} \, ,
\end{eqnarray*}
where $\AGEN{\PW}$ and $\AGEN{\PZ}$ are the fiducial and kinematic acceptances
for $\Wln$ and $\Zll$, respectively, and $\varepsilon_{\PW}$
and $\varepsilon_{\PZ}$ are the selection efficiencies for $\PW$ and $\PZ$ 
signal events in the acceptance.
The uncertainty from $\AGEN{\PW}/\AGEN{\PZ}$ is determined from Monte Carlo 
generator studies to be approximately 1\%.  
\par
The ratio of cross sections for $\PWp$ and $\PWm$ production is 
\begin{eqnarray*}
\RPM = \frac{\SIGBRSHORT{\PWp}}{\SIGBRSHORT{\PWm}} 
= \frac{\NSIG{\PWp}}{\NSIG{\PWm}}\frac{\RHO{\PWm}}{\RHO{\PWp}}\frac{\APRIM{\PWm}}{\APRIM{\PWp}} 
= \frac{\NSIG{\PWp}}{\NSIG{\PWm}}\frac{\EPSB{\PWm}}{\EPSB{\PWp}}\frac{\AGEN{\PWm}}{\AGEN{\PWp}} \, ,
\end{eqnarray*}
where $\AGEN{\PWp}$ and $\AGEN{\PWm}$ are the fiducial and kinematic acceptances
for $\Wpln$ and $\Wmln$, respectively, and $\varepsilon_{\PWp}$
and $\varepsilon_{\PWp}$ are the selection efficiencies for $\PWp$ and $\PWm$ 
signal events in the acceptance.
The uncertainty from ${\AGEN{\PWp}}/{\AGEN{\PWm}}$ is determined from Monte Carlo
generator studies to be approximately 2\%.

\begin{table}[hbt]
\caption[.]{ Summary of the cross section ratio measurements
and their theoretical predictions. \label{tab:res-ratios} }
\begin{center}
\begin{tabular}{|l|c|r|c|}
\hline
\multicolumn{2}{|c|}{Quantity} &  
\multicolumn{1}{c|}{ Ratio } & 
\multicolumn{1}{c|}{NNLO}   \\
\hline
\hline
\multirow{3}{*}{\RWZ} & $\Pe$ & \RESRATWZGHME & \multirow{3}{*}{\THGHMRATIOWZ} \\
                  & $\Pgm$ & \RESRATWZGHMM & \\
                  & $\ell$ & \RESRATWZGHML & \\
\hline
\hline
\multirow{3}{*}{\RPM} & $\Pe$ & \RESRATWWGHME & \multirow{3}{*}{\THGHMRATIOWW} \\
                  & $\Pgm$ & \RESRATWWGHMM & \\
                  & $\ell$ & \RESRATWWGHML & \\
\hline 
\end{tabular}
\end{center}
\end{table}

\par
The measurements of the $\RWZ$ and $\RPM$ cross section ratios are reported 
in Table~\ref{tab:res-ratios}, along with their theoretical predictions.

\par
We also report the cross sections as
measured within the fiducial and kinematic acceptance,
thereby eliminating the PDF uncertainties
from the results.  
In effect, these
uncertainties are transferred to the theoretical predictions,
allowing for a cleaner separation of experimental and
theoretical uncertainties.
For each channel the fiducial and kinematic acceptance 
is defined by the fraction 
of events with lepton $\pt$ greater than $20~\GeV$ after final-state 
QED radiation, and with pseudorapidity in the range $|\eta|<2.5$ for electrons 
and $|\eta|<2.1$ for muons.  

\begin{table}[hbt]
\caption[.]{ Summary of production cross section measurements 
in restricted fiducial and kinematic acceptances. The $\pt$ and $|\eta|$ 
criteria restricting the acceptance for electrons and muons,
and the resulting acceptance values, are also given.
\label{tab:res-restricted-xsections} }
\begin{center}
\begin{tabular}{|l|r|c|c|}
\hline
\multicolumn{1}{|c|}{Channel} &  
\multicolumn{1}{c|}{$\sigma \times {\cal B}$ in acceptance $A$ (nb)} & 
\multicolumn{2}{c|}{$A$} 
\\
\hline
\hline
$\Wen$ 
& \WEISIGBRXGHM 
& \WEIAGEN 
& \\
$\Wpen$ 
& \WEPSIGBRXGHM 
& \WEPAGEN 
& $\pt>20\GeV$  \\
$\Wmen$ 
& \WEMSIGBRXGHM 
& \WEMAGEN 
&       $|\eta|<2.5$   \\
$\Zee$ 
& \ZEESIGBRXGHM 
&  \ZEEAGEN 
& \\
\hline 
\hline
$\Wmn$ 
& \WMISIGBRXGHM 
& \WMIAGEN 
& \\
$\Wpmn$ 
& \WMPSIGBRXGHM 
& \WMPAGEN 
& $\pt>20\GeV$  \\
$\Wmmn$ 
& \WMMSIGBRXGHM 
& \WMMAGEN 
&       $|\eta|<2.1$   \\
$\Zmm$ 
& \ZMMSIGBRXGHM 
&  \ZMMAGEN 
& \\
\hline 
\end{tabular}
\end{center}
\end{table}
\par
The measurements of cross sections in restricted acceptance regions are 
reported in Table~\ref{tab:res-restricted-xsections}, along with 
the acceptance values, computed using the
POWHEG generator, which is complete to the NLO and 
interfaced with PYTHIA for final-state radiation (FSR).
Acceptance values from FEWZ, which is complete
to the NNLO but lacks FSR, are compatible with those from POWHEG.
The quoted errors on the acceptances are due to the PDF uncertainties.
Since the acceptances are different for
electrons and muons, these cross section values cannot
be combined. 
The difference in acceptance for $\PWp$ and $\PWm$,
larger in the electron channel,
is a consequence of the pseudorapidity distributions of $\ell^+$ and 
$\ell^-$ from boson decays, 
which reflect not only the different $x$ distributions of quarks and 
antiquarks in the proton, but also
a distinction between valence and sea quarks at a given $x$ due 
to the V--A interaction. 

\begin{table}
\caption[.]{ \label{tab:RatioCMSTHY}
The ratios 
of the $\PW$ and $\PZ$ cross section times branching ratio measurements
to their theoretical predictions, and of the measured cross section
ratios to their theoretical predictions. 
The uncertainty in the 
integrated luminosity cancels out in the latter ratios.
}
\begin{center}
\begin{tabular}{|c|l|r|c|}
\hline
\multicolumn{2}{|c|}{Quantity} &  
\multicolumn{1}{c|}{ Ratio (CMS/Theory) } & 
\multicolumn{1}{c|}{ Lumi. Uncertainty } \\
\hline
\multirow{4}{*}{$\sigma \times {\cal B}$ } &
$\PW$   & $\RATCMSTHYWI$ & \multirow{4}{*}{$\pm 0.11$}\\
& $\PWp$     & $\RATCMSTHYWP$ & \\
& $\PWm$     & $\RATCMSTHYWM$ & \\
& $\PZ$       & $\RATCMSTHYZ$  & \\
\hline
\hline
\multicolumn{2}{|c|}{\RWZ}     & $\RATCMSTHYWZ$ & \multirow{2}{*}{nil} \\
\multicolumn{2}{|c|}{\RPM}     & $\RATCMSTHYWW$ & \\
\hline
\end{tabular}
\end{center}
\end{table}

\begin{figure}
\begin{center}
  \includegraphics[width=0.80\textwidth]{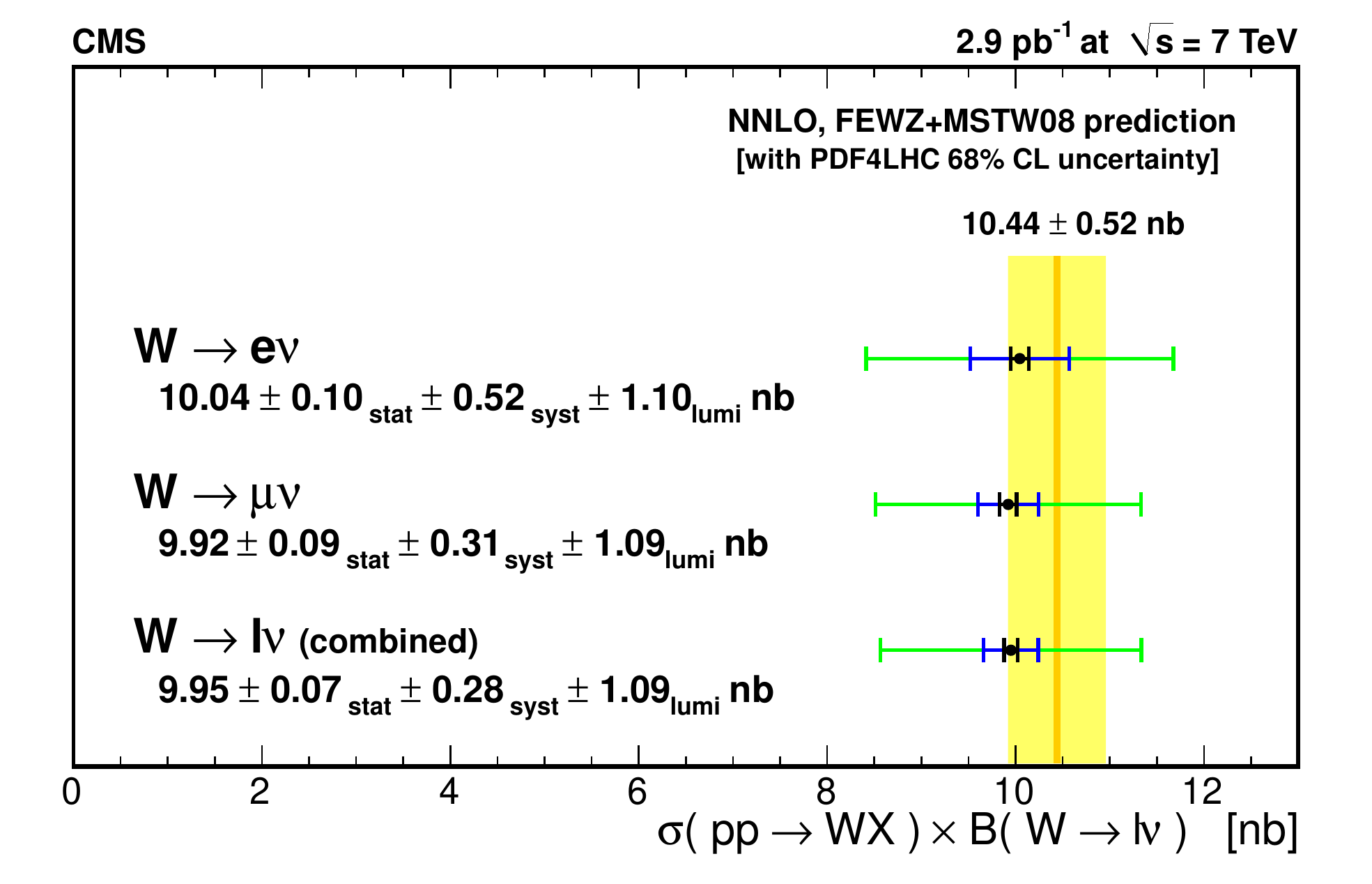}
\caption[.]{\label{fig:W_LEPstylePlots}
Summary of the $\PW$ boson production cross section times branching ratio
measurements.}
\end{center}
\end{figure}

\begin{figure}
\begin{center}
  \includegraphics[width=0.80\textwidth]{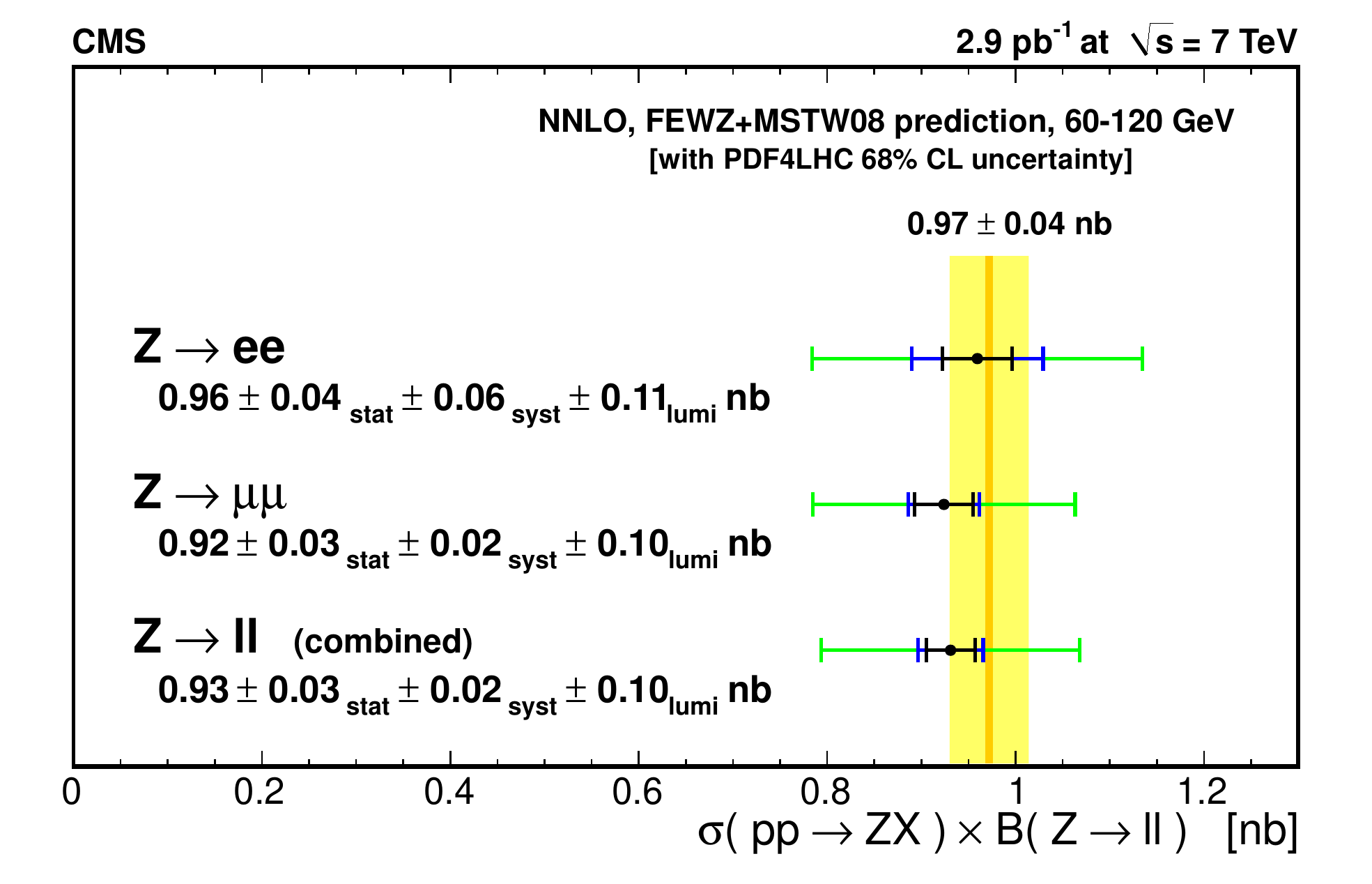}
\caption[.]{\label{fig:Z_LEPstylePlots}
Summary of the $\PZ$ boson production cross section times branching ratio
measurements.}
\end{center}
\end{figure}

\begin{figure}
\begin{center}
  \includegraphics[width=0.80\textwidth]{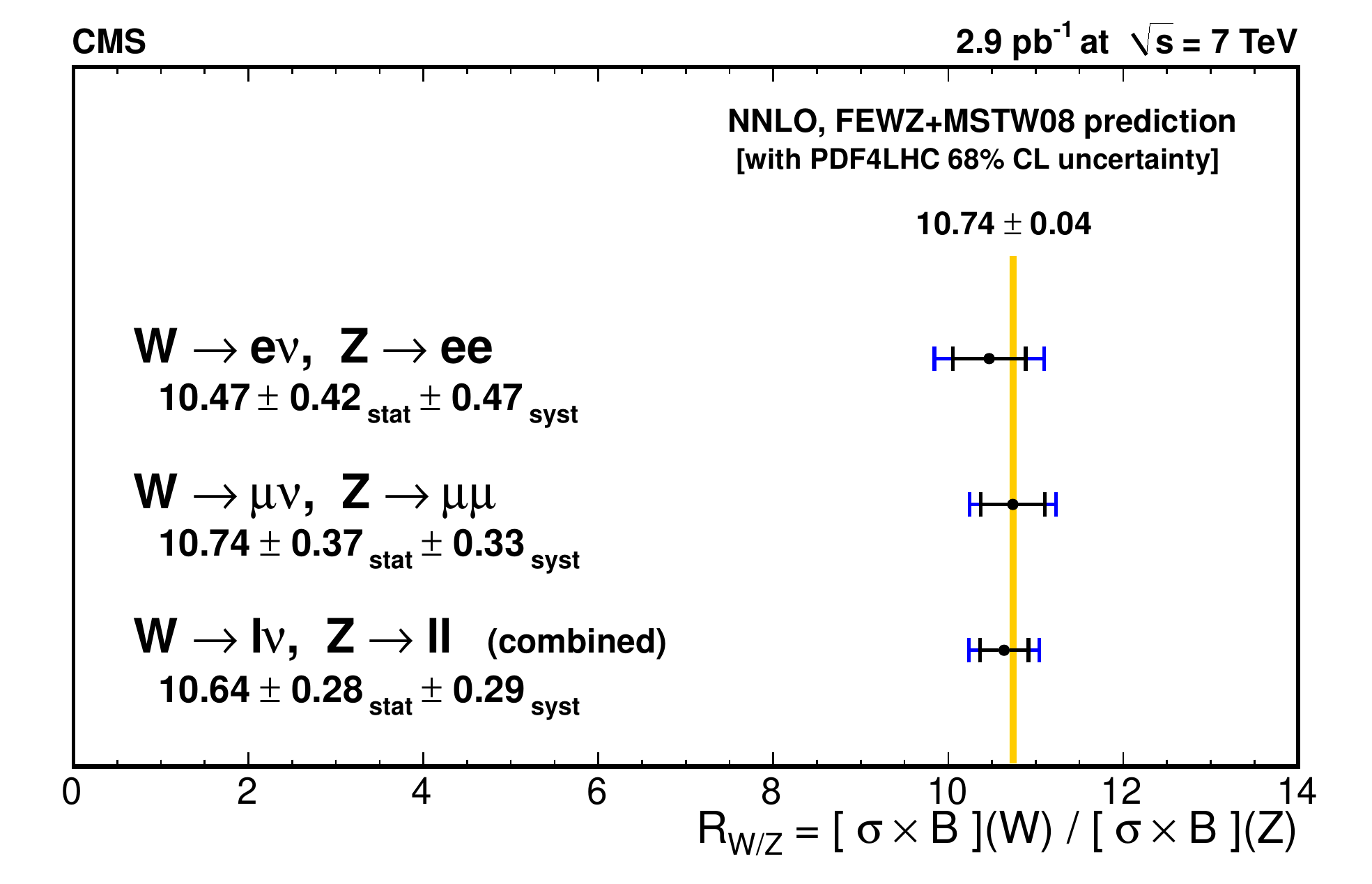}
\caption[.]{\label{fig:WZ_LEPstylePlots}
Summary of the $\RWZ$ cross section ratio measurements.}
\end{center}
\end{figure}

\begin{figure}
\begin{center}
  \includegraphics[width=0.80\textwidth]{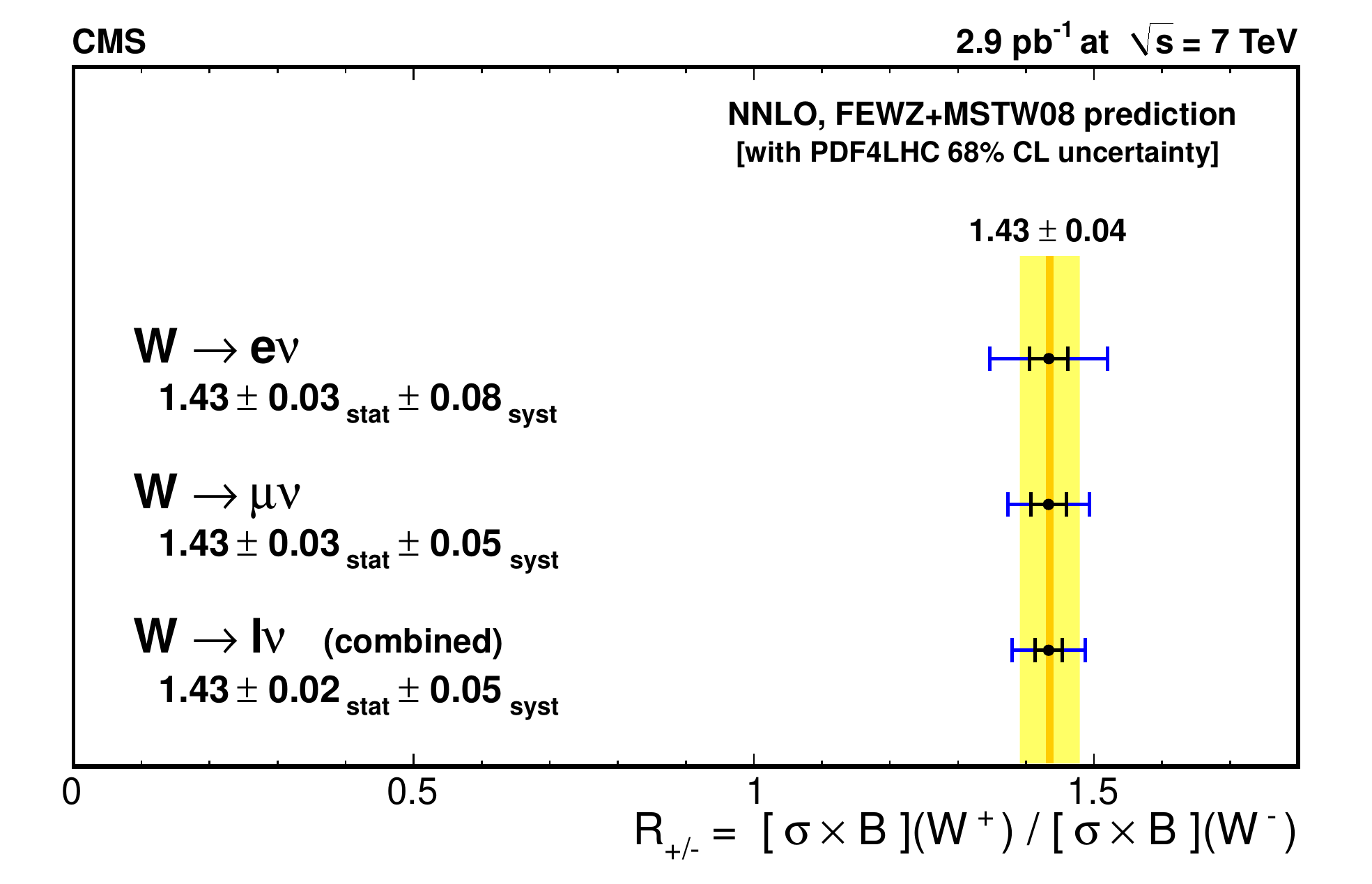}
\caption[.]{\label{fig:WPM_LEPstylePlots}
Summary of the $\RPM$ cross section ratio measurements.}
\end{center}
\end{figure}

\begin{figure}
\begin{center}
  \includegraphics[width=0.8\textwidth]{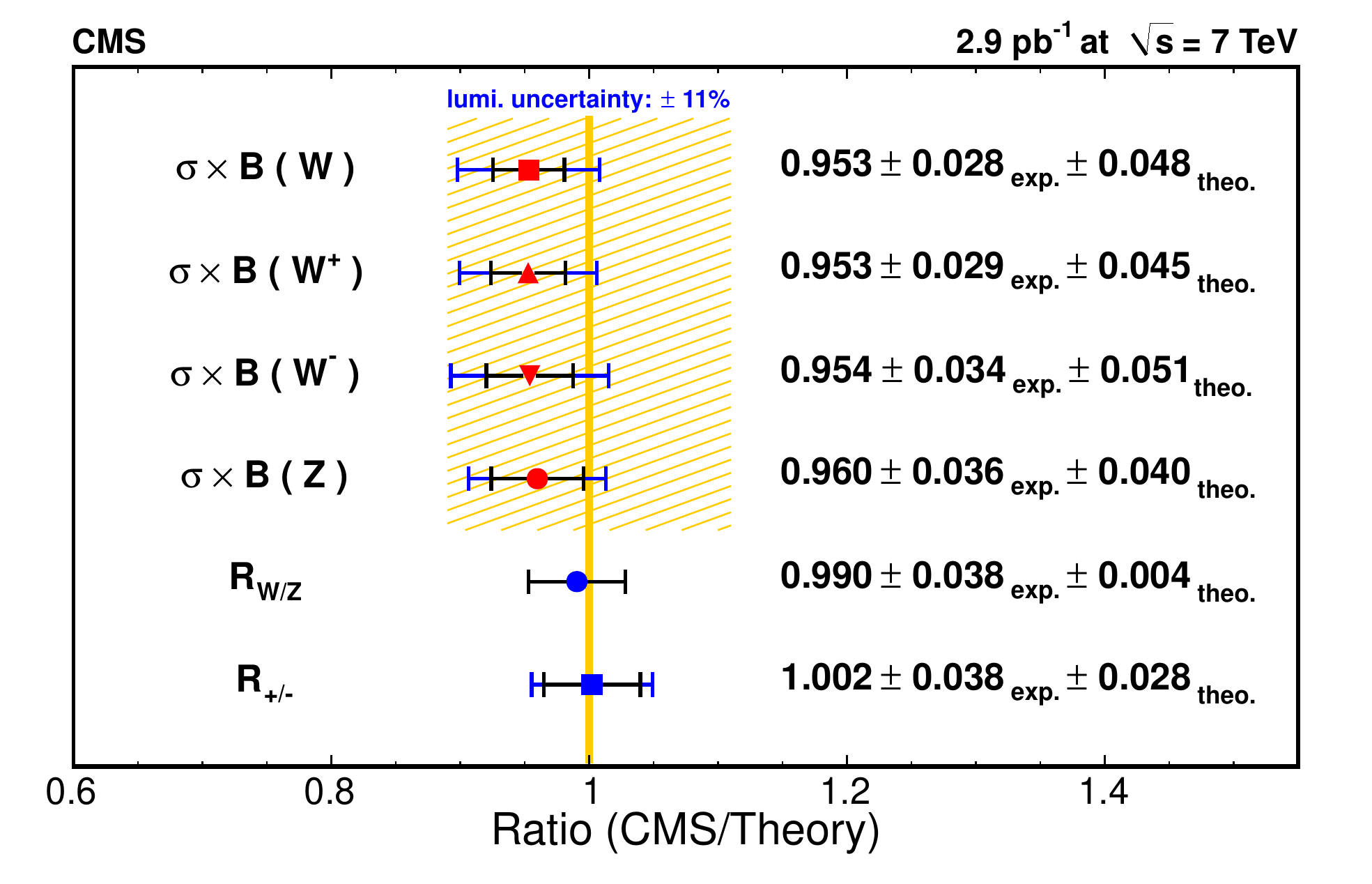}
\caption[.]{\label{fig:RatioCMSTHY}
Summary of the ratios of the CMS measurements to their theoretical 
predictions.
The luminosity uncertainty ($\pm 11\%$), which 
affects only the cross section times branching ratio measurements, 
is represented by 
a shaded area.}
\end{center}
\end{figure}

Summaries of the measurements are given 
in Figs.~\ref{fig:W_LEPstylePlots},~\ref{fig:Z_LEPstylePlots},~\ref{fig:WZ_LEPstylePlots}, and~\ref{fig:WPM_LEPstylePlots},
illustrating  the consistency of the measurements in the electron
and muon channels, as well as the confirmation
of theoretical
predictions computed at the NNLO in QCD with state-of-the-art PDF sets. 
For each reported measurement, the statistical error 
is represented in black and
the total experimental uncertainty, obtained by adding in quadrature the
statistical and systematic uncertainties, in dark blue.
For the cross section measurements,  the luminosity
uncertainty is added linearly to the experimental uncertainty, 
and is represented in green. The dark-yellow  
vertical line represents the 
theoretical prediction, and the light-yellow vertical band is 
the theoretical uncertainty, interpreted as a 68\% confidence interval, 
as described above. 

\begin{figure} [htb]
\begin{center}
\includegraphics[width=1.0\textwidth,angle=0]{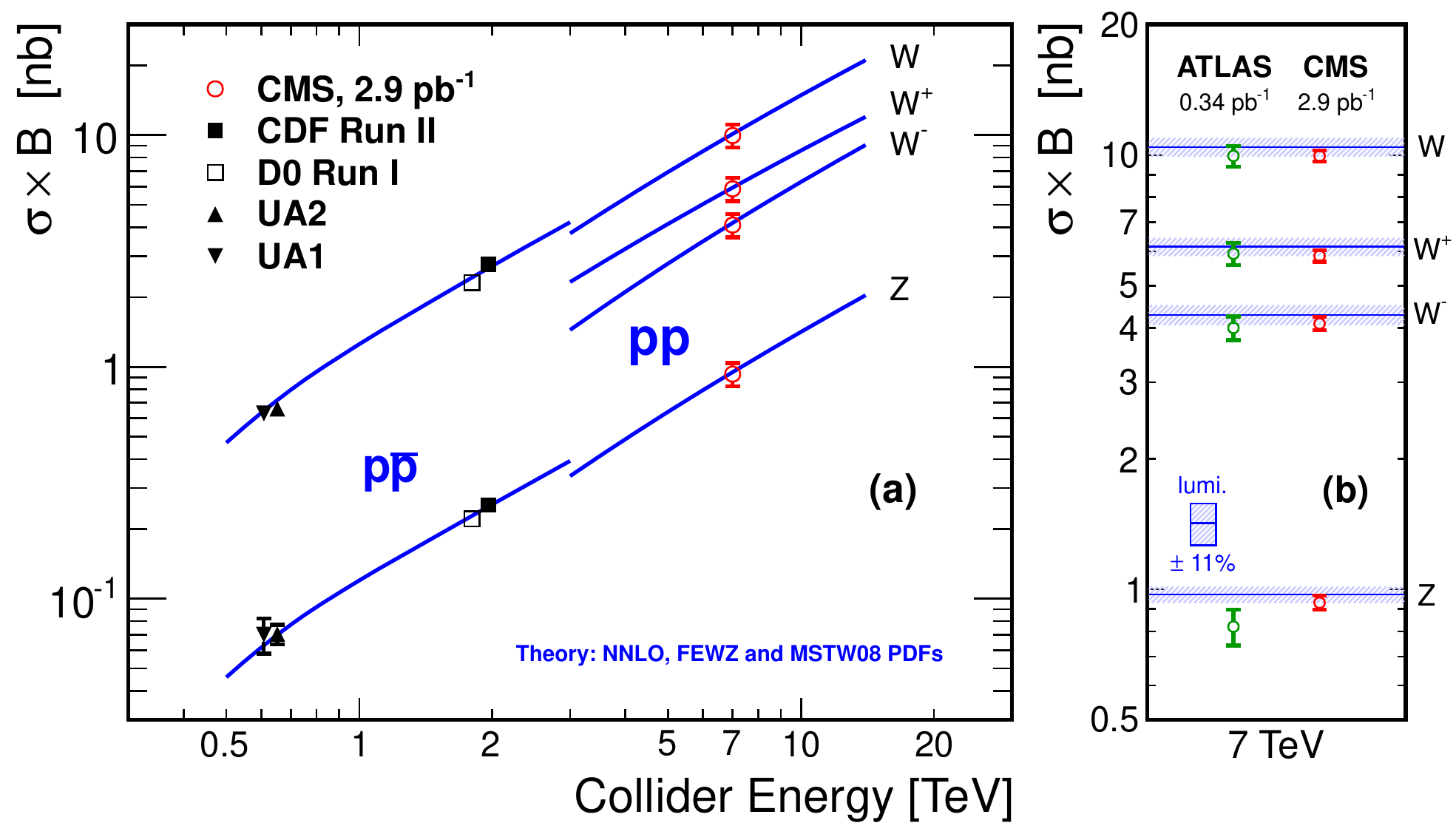}
\caption[.]{\label{fig:WZsigmas}
(a) Measurements of inclusive $\PW$ and $\PZ$ production 
cross sections times branching ratio as a function
of center-of-mass energy 
for CMS and  experiments
at lower-energy colliders. 
The lines are the NNLO theory predictions.
(b) Comparison of the ATLAS and CMS $\PW$ and $\PZ$ production
cross sections times branching ratios. The error bars are the 
statistical and systematic uncertainties added in quadrature,
except for the uncertainty on the integrated luminosity, 
whose value is shown separately as a band.
}
\end{center}
\end{figure}

\par
The agreement of theoretical predictions with our measurements 
is quantified
in Table~\ref{tab:RatioCMSTHY} and illustrated 
in Fig.~\ref{fig:RatioCMSTHY}. 
There, the experimental uncertainty ("exp") is computed as the sum
in quadrature of the statistical uncertainty and the
systematic uncertainties aside from the theoretical
uncertainties associated with the acceptance.
The theoretical uncertainty ("theo") is computed by adding
in quadrature the variations of the central value when
the renormalization scale is varied, and the PDF uncertainty.  
Figure~\ref{fig:WZsigmas}~(a) represents the CMS measurements together 
with measurements at lower-energy hadron colliders.
The increase of the $\PW$ and $\PZ$ cross sections with energy is confirmed. 
Fig.~\ref{fig:WZsigmas}~(b) shows the good agreement  
between CMS and ATLAS measurements in $\pp$ collisions 
at $\rts=7~\TeV$.
\section{Conclusions}
\label{sec:conclusions}
\par
We have performed measurements of 
inclusive $\PW$ and $\PZ$ boson production cross sections 
in $\pp$ collisions at $\rts=7~\TeV$
using $\LUMI$ of data recorded by the CMS detector at the LHC.
We find internal consistency between measurements in 
the electron and muon channels
and report their combination.
We also report ratios of 
$\PW$ 
to
$\PZ$ 
and
$\PWp$ 
to
$\PWm$ 
production 
cross sections. 
The theoretical predictions agree with our measurements, 
as illustrated in Fig.~\ref{fig:RatioCMSTHY}.
\par
Aside from the luminosity uncertainty, which cancels 
in the ratios, the systematic uncertainties 
are comparable to the statistical ones in our measurements.
The experimental uncertainties are smaller than those on the theoretical
predictions; they are typically less than 4\%.  This suggests
that the inclusive measurements of $\PW$ and $\PZ$ cross sections
can potentially be used to normalize the LHC luminosity at the 5\% level 
or better~\cite{Dittmar:1997md}.
As 
most of
the systematic uncertainties 
are
statistical in nature,  they will decrease with larger data
samples, and also benefit from an improved understanding of the CMS detector.

\section*{Acknowledgements} 

\hyphenation{Bundes-ministerium For-schungs-gemeinschaft Forschungs-zentren}\hyphenation{Helmholtz-Gemeinschaft Deutscher Forschung-szentren} We wish to congratulate our colleagues in the CERN accelerator departments for the excellent performance of the LHC machine. We thank the technical and administrative staff at CERN and other CMS institutes. This work was supported by the Austrian Federal Ministry of Science and Research; the Belgium Fonds de la Recherche Scientifique, and Fonds voor Wetenschappelijk Onderzoek; the Brazilian Funding Agencies (CNPq, CAPES, FAPERJ, and FAPESP); the Bulgarian Ministry of Education and Science; CERN; the Chinese Academy of Sciences, Ministry of Science and Technology, and National Natural Science Foundation of China; the Colombian Funding Agency (COLCIENCIAS); the Croatian Ministry of Science, Education and Sport; the Research Promotion Foundation, Cyprus; the Estonian Academy of Sciences and NICPB; the Academy of Finland, Finnish Ministry of Education, and Helsinki Institute of Physics; the Institut National de Physique Nucl\'eaire et de Physique des Particules~/~CNRS, and Commissariat \`a l'\'Energie Atomique, France; the Bundesministerium f\"ur Bildung und Forschung, Deutsche Forschungsgemeinschaft, and Helmholtz-Gemeinschaft Deutscher Forschungszentren, Germany; the General Secretariat for Research and Technology, Greece; the National Scientific Research Foundation, and National Office for Research and Technology, Hungary; the Department of Atomic Energy, and Department of Science and Technology, India; the Institute for Studies in Theoretical Physics and Mathematics, Iran; the Science Foundation, Ireland; the Istituto Nazionale di Fisica Nucleare, Italy; the Korean Ministry of Education, Science and Technology and the World Class University program of NRF, Korea; the Lithuanian Academy of Sciences; the Mexican Funding Agencies (CINVESTAV, CONACYT, SEP, and UASLP-FAI); the Pakistan Atomic Energy Commission; the State Commission for Scientific Research, Poland; the Funda\c{c}\~ao para a Ci\^encia e a Tecnologia, Portugal; JINR (Armenia, Belarus, Georgia, Ukraine, Uzbekistan); the Ministry of Science and Technologies of the Russian Federation, and Russian Ministry of Atomic Energy; the Ministry of Science and Technological Development of Serbia; the Ministerio de Ciencia e Innovaci\'on, and Programa Consolider-Ingenio 2010, Spain; the Swiss Funding Agencies (ETH Board, ETH Zurich, PSI, SNF, UniZH, Canton Zurich, and SER); the National Science Council, Taipei; the Scientific and Technical Research Council of Turkey, and Turkish Atomic Energy Authority; the Science and Technology Facilities Council, UK; the US Department of Energy, and the US National Science Foundation.

Individuals have received support from the Marie-Curie programme and the European Research Council (European Union); the Leventis Foundation; the A. P. Sloan Foundation; the Alexander von Humboldt Foundation; the Associazione per lo Sviluppo Scientifico e Tecnologico del Piemonte (Italy); the Belgian Federal Science Policy Office; the Fonds pour la Formation \`a la Recherche dans l'Industrie et dans l'Agriculture (FRIA-Belgium); and the Agentschap voor Innovatie door Wetenschap en Technologie (IWT-Belgium).

\bibliography{auto_generated}   

\cleardoublepage\appendix\section{The CMS Collaboration \label{app:collab}}\begin{sloppypar}\hyphenpenalty=5000\widowpenalty=500\clubpenalty=5000\textbf{Yerevan Physics Institute,  Yerevan,  Armenia}\\*[0pt]
V.~Khachatryan, A.M.~Sirunyan, A.~Tumasyan
\vskip\cmsinstskip
\textbf{Institut f\"{u}r Hochenergiephysik der OeAW,  Wien,  Austria}\\*[0pt]
W.~Adam, T.~Bergauer, M.~Dragicevic, J.~Er\"{o}, C.~Fabjan, M.~Friedl, R.~Fr\"{u}hwirth, V.M.~Ghete, J.~Hammer\cmsAuthorMark{1}, S.~H\"{a}nsel, C.~Hartl, M.~Hoch, N.~H\"{o}rmann, J.~Hrubec, M.~Jeitler, G.~Kasieczka, W.~Kiesenhofer, M.~Krammer, D.~Liko, I.~Mikulec, M.~Pernicka, H.~Rohringer, R.~Sch\"{o}fbeck, J.~Strauss, A.~Taurok, F.~Teischinger, W.~Waltenberger, G.~Walzel, E.~Widl, C.-E.~Wulz
% ...
\vskip\cmsinstskip
% ...
1:~~Also at CERN, European Organization for Nuclear Research, Geneva, Switzerland\\
% ...

\end{sloppypar}
\end{document}